\documentclass[journal]{IEEEtran}
\usepackage[latin9]{inputenc}
\usepackage{color}
\usepackage{amsmath}
\usepackage{amsthm}
\usepackage{amssymb}
\usepackage{graphicx}

\makeatletter

\DeclareTextSymbolDefault{\textquotedbl}{T1}
\newcommand{\lyxdot}{.}

\theoremstyle{plain}
\newtheorem{thm}{\protect\theoremname}
\theoremstyle{plain}
\newtheorem{prop}[thm]{\protect\propositionname}
\theoremstyle{remark}
\newtheorem{rem}[thm]{\protect\remarkname}
\theoremstyle{plain}
\newtheorem{cor}[thm]{\protect\corollaryname}

\usepackage[caption=false,font=footnotesize]{subfig}

\theoremstyle{plain}

\newtheorem*{ass*}{Assumption}
\allowdisplaybreaks
\usepackage{flushend}

\@ifundefined{showcaptionsetup}{}{%
 \PassOptionsToPackage{caption=false}{subfig}}
\usepackage{subfig}
\makeatother

\providecommand{\corollaryname}{Corollary}
\providecommand{\propositionname}{Proposition}
\providecommand{\remarkname}{Remark}
\providecommand{\theoremname}{Theorem}

\begin{document}
\title{Non-Coherent Joint Transmission in Poisson Cellular Networks Under
Pilot Contamination}
\author{Stelios~Stefanatos and Gerhard~Wunder~\thanks{This work has been performed in the framework of the Horizon 2020
project ONE5G (ICT-760809) receiving funds from the European Union.
The authors would like to acknowledge the contributions of their colleagues
in the project, although the views expressed in this contribution
are those of the authors and do not necessarily represent the project.
The work of G. Wunder was also supported by DFG grants WU 598/7-1
and WU 598/8-1 (DFG Priority Program on Compressed Sensing).\protect \\
S. Stefanatos and G. Wunder are with the Department of Mathematics
and Computer Science, Freie Universit\"at Berlin, 14195, Berlin,
Germany, e-mail: \{stelios.stefanatos, g.wunder\}@fu-berlin.de.}}
\maketitle
\begin{abstract}
This paper investigates the performance of downlink cellular networks
with non-coherent joint (mutlipoint) transmissions and practical channel
estimation. Under a stochastic geometry framework, the spatial average
signal-to-noise-ratio (SNR) is characterized, taking into account
the effect of channel estimation error due to pilot contamination.
A simple, easy to compute SNR expression is obtained under the assumption
of randomly generated pilot sequences and minimal prior information
about the channels and positions of access points (APs). This SNR
expression allows for the efficient joint optimization of critical
system design parameters such as number of cooperating APs and training
overhead. Among others, it is shown that multipoint transmissions
are preferable to conventional (non-cooperative) cellular operation
under certain operational conditions. Furthermore, analytical insights
are obtained regarding (a) the minimum training overhead required
to achieve a given SNR degradation compared to the perfect channel
estimation case and (b) the optimal number of cooperating APs when
an arbitrarily large training overhead can be afforded. For the latter,
in particular, a phase transition phenomenon is identified, where
the optimal number of cooperating APs is either finite or infinite,
depending on whether the path loss factor is less or equal than a
certain value, respectively.
\end{abstract}


\IEEEpeerreviewmaketitle{}

\section{Introduction}

The densification of the cellular network infrastructure results in
any user equipment (UE) in the system likely to be positioned in the
proximity of multiple access points (APs), which naturally promotes
mutlipoint transmission techniques \cite{network densification}.
Considering the downlink of a frequency division duplex (FDD) system,
non-coherent joint transmission (NCJT), where the serving APs transmit
the same signal without any precoding \cite{NCJT better than CJT conf},
is particularly attractive as it eliminates the need for channel state
information (CSI) at the APs and, in turn, the associated feedback
overhead. However, downlink CSI is still required at the UE side in
order to (coherently) decode the received signal. Towards estimating
the (effective) downlink channel, a training phase preceding the data
transmission phase is commonly employed, during which APs transmit
known pilot (training) sequences.

Of course, the quality of the channel estimate will have an impact
on the performance, since the residual channel estimation error effectively
increases the noise level \cite{Hassibi}. It is therefore important
to design the training phase such that channel estimates of sufficient
accuracy can be obtained. For the dense cellular setting, this design
should not only have to take into account the effects of the ambient
additive noise at the UE side but also the effect of interference
from pilot transmissions by nearby APs not serving the UE under consideration.
The latter effect is particularly dominant in the massive MIMO setting,
commonly referred to as pilot contamination \cite{Marzetta}. Nevertheless,
pilot contamination is present in every network with non-orthogonal
pilot transmissions, even with single antenna transceivers. In addition,
it is desirable to design the training phase such that the overhead
associated with the training phase itself, as well as the acquisition
of any prior channel information employed at the UE side, is minimized.
This is especially important in scenarios with small channel coherence
intervals (mobility). Clearly, towards achieving these challenging
design goals, a tractable characterization of the system performance
that captures the effects of both channel estimation error and overhead
is desirable.

\subsection{Previous Work}

The theoretical benefits of multipoint transmissions under perfect
CSI at the UE(s) and, possibly, at the AP(s) side are well documented
in the information-theoretic literature \cite{Gesbert}, under simplistic
assumptions on the AP positions. However, as the network infrastructure
becomes more dense, the AP positions exhibit a random behavior, which
is expected to have impact on the system performance.

Towards incorporating the effects of large scale fading and randomness
of AP positions, stochastic geometry (SG) has become a popular modeling
tool that allows for tractable performance characterization of cellular
networks \cite{Elsawy tutorial}. Under a perfect CSI assumption,
SG has been successfully employed to obtain simple closed form expressions
for the signal-to-noise-ratio ($\mathsf{SNR}$) coverage probabilities
of the conventional (non-cooperative) cellular network (e.g., see
\cite{Andrews tractable} for downlink and \cite{elsawy uplink} for
uplink analysis). However, extension of this approach to multipoint
transmissions and/or imperfect channel estimation faces significant
analytical complications.

Focusing on NCJT systems that are of interest in this paper, \cite{Nigam}
provides the exact expression for the $\mathsf{SNR}$ coverage probability
under perfect CSI. This expression requires the numerical computation
of multiple integrals, restricting its practical application to clusters
of only up to $3$ cooperative APs. Under the same setting, \cite{Tanbourgi}
provides more computationally friendly coverage probability expressions
based on various approximations, however, the expressions are still
quite complicated to offer significant analytical insights. Similar
remarks hold also for the NCJT performance analysis under perfect
CSI presented in \cite{Garcia}.

Regarding the effects of channel estimation error, a surprisingly
limited number of works under the SG framework exist. Most of these
works consider the non-cooperative massive MIMO setting \cite{Kountouris,Bai Heath,Dhillon massive MIMO},
which allows certain analytical simplifications due to the channel
hardening effect present there \cite{cell free vs small cells}. However,
employing the same analysis in a non-massive MIMO setting results
in a crude approximation as channel hardening no longer holds \cite{Bjornson channel hardening}.
In addition, \cite{Kountouris,Bai Heath} consider a Bayesian channel
estimation with knowledge of every AP-UE distance in the system. This
assumption implies a signaling overhead for obtaining this information
whose effect is not taken into account in the analysis. The only SG
work in a non-massive MIMO setting that explicitly focuses on imperfect
CSI effects appears to be \cite{Heath channel estimation conf}. Analysis
is limited to a non-cooperative transmission scenario, assuming perfect
knowledge of the serving AP distance, and the resulting coverage probability
expression provides limited analytical insights. The cost of imperfect
CSI in NCJT is briefly treated also in \cite{Tanbourgi}, however,
assuming the availability of orthogonal (i.e., non-interfering) pilot
sequences.

\subsection{Contributions}

This paper attempts to provide a tractable system performance characterization
under NCJT and practical channel estimation for a non-massive MIMO
setting using the SG modeling framework. The ultimate goal is to obtain
a simple metric that can be used to not only numerically optimize
important system design parameters such as number of cooperating APs
and training overhead, but also provide analytical insights on how
their optimal values depend on operational conditions such as path
loss factor, average transmit power and ambient noise level. In particular,
the following paper contributions can be identified.
\begin{itemize}
\item Towards minimum signaling overhead, channel estimation under minimal
assumptions on the prior information about AP channels and positions
is considered. This renders the channels to be estimated non Gaussian
and specification of the channel estimator is not straightforward.
In addition, pilot sequences of small length (overhead) are considered,
which are necessarily non-orthogonal and hence introduce pilot contamination
effects. By considering randomly generated AP pilot sequences, which
requires no coordination/overhead for pilot assignment, a closed form
expression for the linear minimum mean square error (LMMSE) estimator
as well as its mean square error is provided.
\item Instead of the standard, $\mathsf{SNR}$ coverage probability considered
in previous works, the less informative but much more tractable spatial
average $\mathsf{SNR}$ is considered as the performance metric of
the system. The $\mathsf{SNR}$ computation takes explicitly the effect
of residual channel estimation error into account and the resulting
expression highlights the dependence of the $\mathsf{SNR}$ on channel
estimation accuracy, number of cooperative APs and noise level. It
is shown that, even though the conventional, non-cooperative cellular
network is sufficient in terms of $\mathsf{SNR}$ under perfect CSI,
a cooperative cluster of $2$ or more APs can provide better performance
under imperfect channel estimation and operational conditions where
the impact of pilot contamination is greater than that of additive
noise in the training phase.
\item The $\mathsf{SNR}$ expression is numerically optimized with respect
to (w.r.t.) critical system design parameters such as number of cooperative
APs and pilot sequence length (training overhead), where it is shown
that cooperation of multiple APs, potentially much greater than $2$,
is mostly beneficial under propagation conditions with a large path
loss factor. In addition, interesting analytical insights are obtained
by manipulation of the $\mathsf{SNR}$ expression regarding (a) the
minimum training overhead required to achieve a given $\mathsf{SNR}$
degradation compared to the perfect CSI case and (b) the optimal number
of cooperating APs when an arbitrarily large training overhead can
be afforded. For the latter, in particular, a phase transition phenomenon
is identified, where the optimal number of cooperating APs is either
finite or infinite, depending on the path loss factor being less or
equal than a certain value, respectively.
\end{itemize}

\subsection{Notation}

Boldface lower (upper) case letters denote column vectors (matrices).
The transposition and Hermitian transposition are denoted by $(\cdot)^{T}$
and $(\cdot)^{H}$, respectively. A scalar random variable will be
said to be distributed as $\mathcal{CN}(0,\sigma^{2})$ if it is distributed
as a circularly-symmetric complex Gaussian random variable of zero
mean and variance $\sigma^{2}$. The expectation operator is denoted
by $\mathbb{E}(\cdot)$. The Euclidean norm of a vector $\mathbf{x}$
is denoted by $\|\mathbf{x}\|$. The $N\times N$ identity matrix
is denoted by $\mathbf{I}_{N}$ and $\mathbf{1}$ denotes the all-ones
column vector whose dimension will be clear from context. For two
positive-valued functions $f,$ $g$, the notation $f(x)=\mathcal{O}(g(x)),x\rightarrow\infty(x\rightarrow0)$,
is used to represent the condition $f(x)\leq Mg(x),x>x_{0}(x<x_{0})$,
for a sufficiently large (small) $x_{0}>0$ and some constant $M>0$
that is independent of $x$.

\section{System Model}

The downlink of a dense cellular network covering a wide geographical
area is considered. The locations on the plane of the APs are modeled
as a realization of a homogeneous Poisson point process (HPPP) $\Phi\subset\mathbb{R}^{2}$
of density $\lambda>0$ (average number of APs per unit area) \cite{Andrews primer PPP}.
All APs and UEs in the system are equipped with a single antenna.
By the stationarity of the HPPP \cite{Haenggi Ganti book}, a typical
UE located at the origin of the plane will be considered in the following.

The baseband-equivalent flat fading channel $h_{\mathbf{x}}$, corresponding
to the downlink between an AP located at $\mathbf{x}\in\Phi$ and
the typical UE, follows the standard model \cite{Haenggi Ganti book}
\begin{equation}
h_{\mathbf{x}}=c_{\mathbf{x}}\sqrt{\ell(\|\mathbf{x}\|)},\label{eq:channel model}
\end{equation}
where $c_{\mathbf{x}}\in\mathbb{C}$ represents the small scale fading
and $\ell:[0,\infty)\rightarrow(0,1]$ is a path loss function governing
the large scale fading. The variables $\{c_{\mathbf{x}}\}_{\mathbf{x}\in\Phi}$
are assumed to be i.i.d. as $\mathcal{CN}(0,1)$ (Rayleigh fading
model), while the large scale fading is modeled as \cite{Baccelli}
\begin{equation}
\ell(r)\triangleq r_{0}^{\alpha}\left(\max(r_{0},r)\right)^{-\alpha},r\geq0,\label{eq:large scale fading law}
\end{equation}
where $\alpha>2$ is the path loss factor and $r_{0}>0$ is the reference
distance, i.e., $\ell(r)=1$, for $r\leq r_{0}$.

\begin{figure}[t]
\noindent \centering{}\includegraphics[scale=0.4]{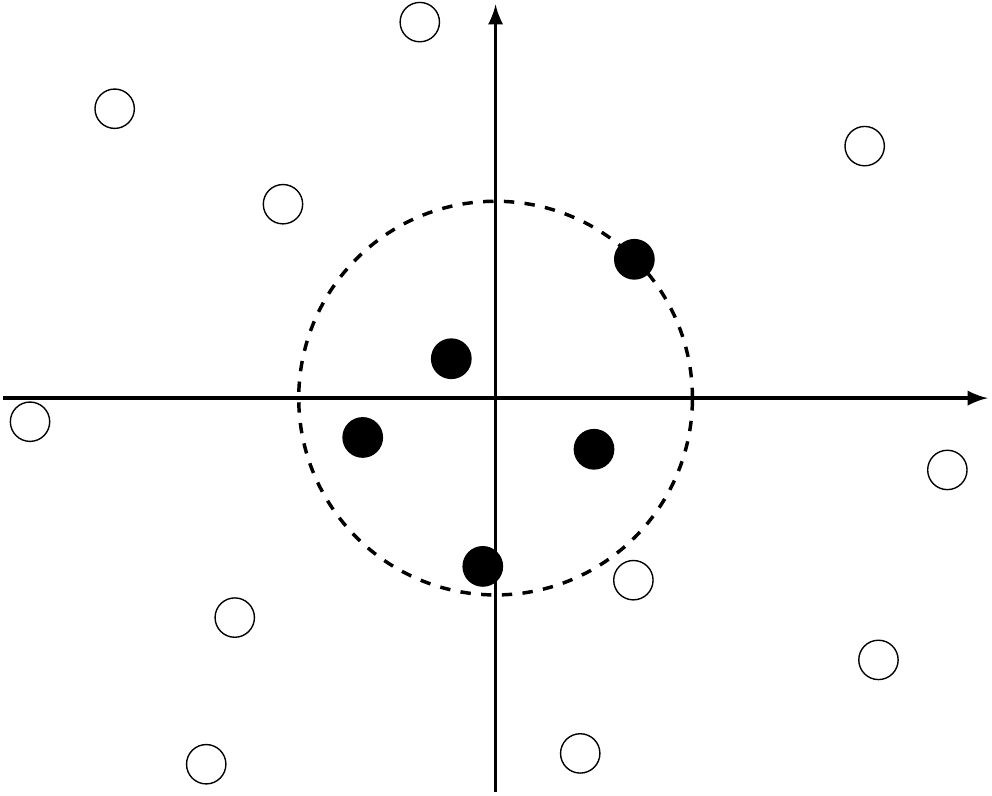}\caption{\label{fig: system model}An AP cluster serving the typical UE at
the origin consisting of $N_{a}=5$ APs whose positions are depicted
as filled circles. During data transmission to the typical UE, APs
outside the cluster, whose positions are depicted as open circles,
are not active.}
\end{figure}

For downlink communications, a cooperative mutlipoint transmission
scheme based on user-cetric adaptive clustering is employed \cite{Gesbert user centric}.
In particular, the typical UE is served by its $N_{a}$ closest in
distance APs, with $N_{a}$ a design parameter that depends in principle
on the operational conditions. An example of an AP distribution realization
and cluster configuration is shown in Fig. \ref{fig: system model}.
The serving cluster employs NCJT, i.e., all APs in the cluster simply
transmit the same signal without any (joint) precoding, which has
the benefit the the APs require no CSI \cite{Nigam,Tanbourgi}. In
addition, it is assumed that there are no interfering transmissions
from out-of-cluster APs during the data transmission phase for the
typical UE, by means of system-wide UE scheduling and/or resource
allocation. The corresponding system performance serves as an upper
bound when interference exists during data transmission.

With all cluster APs transmitting with the same average power, the
received signal at the UE side during data transmission and after
normalization with the average transmit power equals \cite{Nigam}
\begin{align}
y_{d} & =d\sum_{\mathbf{x}\in\mathcal{C}}h_{\mathbf{x}}+w_{d}\nonumber \\
 & =d\mathbf{1}^{T}\mathbf{h}_{\mathcal{C}}+w_{d},\label{eq:data I/O equation}
\end{align}
where $d\in\mathbb{C}$ is a data symbol of zero mean and variance
one, $\mathcal{C}\subseteq\Phi$ is the set of locations of the $N_{a}$
cluster APs,\footnote{$N_{a}$ can become infinite by considering $\mathcal{C}=\Phi$.}
$\mathbf{h}_{\mathcal{C}}\in\mathbb{C}^{N_{p}}$ is a vector with
elements $\{h_{\mathbf{x}}\}_{\mathbf{x}\in\mathcal{C}}$ in some
arbitrary order, and $w_{d}\in\mathbb{C}$ is additive noise distributed
as $\mathcal{CN}(0,\sigma_{w}^{2})$ with $\sigma_{w}^{2}$ representing
the ratio of the additive noise variance to the average transmit power.

For coherent processing of the received signal, the typical UE employs
an estimate of the effective NCJT channel $\mathbf{1}^{T}\mathbf{h}_{\mathcal{C}}$
appearing in (\ref{eq:data I/O equation}), which is obtained by means
of a training phase preceding the data transmission phase. During
the training phase, all APs in the system transmit their pilot sequences
of length $N_{p}$ symbols (channel uses) so that all UEs in the system
are able to simultaneously obtain an estimate of their NCJT channels.
Assuming, without loss of generality, that all APs in the system transmit
during the training phase with the same power as during the data transmission
phase (when active), the typical UE observes the $N_{p}$-dimensional
transmit-power-normalized training signal 
\begin{align}
\mathbf{y}_{p} & =\sum_{\mathbf{x}\in\Phi}h_{\mathbf{x}}\mathbf{p}_{\mathbf{x}}+\mathbf{w}_{p}\nonumber \\
 & =\mathbf{P}_{\mathcal{C}}\mathbf{h}_{\mathcal{C}}+\mathbf{P}_{\mathcal{\bar{C}}}\mathbf{h}_{\mathcal{\bar{C}}}+\mathbf{w}_{p},\label{eq: received pilot}
\end{align}
where $\mathbf{p}_{\mathbf{x}}\in\mathbb{C}^{N_{p}}$ is the pilot
sequence of the AP located at $\mathbf{x}\in\Phi$, $\mathbf{w}_{p}\in\mathbb{C}^{N_{p}}$
is an additive noise vector whose elements are i.i.d. as $\mathcal{CN}(0,\sigma_{w}^{2})$,
$\mathbf{P}_{\mathcal{C}}\in\mathbb{C}^{N_{p}\times N_{a}}$ is a
matrix with columns $\{\mathbf{p}_{\mathbf{x}}\}_{\mathbf{x}\in\mathcal{C}}$
ordered in accordance to the AP ordering in $\mathbf{h}_{\mathcal{C}}$,
and $\mathbf{h}_{\mathcal{\bar{C}}}$, $\mathbf{P}_{\mathcal{\bar{C}}}$
are similarly defined to $\mathbf{h}_{\mathcal{C}}$, $\mathbf{P}_{\mathcal{C}}$,
for the channels and pilot sequences of out-of-cluster APs, respectively.\footnote{Note that $\mathbf{h}_{\mathcal{\bar{C}}}$, $\mathbf{P}_{\mathcal{\bar{C}}}$
have an infinite number of elements and columns, respectively.}

Ideally, the pilot sequences $\{\mathbf{p}_{\mathbf{x}}\}_{\mathbf{x}\in\Phi}$
should be jointly optimized according to the AP and UE locations \cite{locally orthoganal graph theory,graph pilots 2}.
However, this approach requires computationally intensive solvers
as well as additional signaling overhead to obtain location information.
Towards avoiding this, and inspired by the line of works on code-division
multiple access with random spreading sequences \cite{Verdu CDMA,Tse CDMA},
it is assumed in this paper that the AP pilot sequences are generated
randomly and independently, with elements that are i.i.d. as $\mathcal{CN}(0,1)$.

Note that this approach guarantees with probability $1$ (w.p. $1$)
that the pilot sequences are unique, in contrast to the common assumption
of all APs in the system sharing a limited set of pilot sequences
\cite{Marzetta}, which inherently results in (a) channel identifiability
issues due to $\mathbf{P}_{\mathcal{C}}$ not being full column rank
when two or more cluster APs are assigned the same sequence, and (b)
pilot contamination issues due to out-of-cluster APs having the same
training sequences as cluster APs. Even though with the random pilot
assignment there are no identifiability issues since $\mathbf{P}_{\mathcal{C}}$
is full rank w.p. $1$, pilot contamination is still present, since,
w.p. $1$, it holds $\mathbf{p}_{\mathbf{x}}^{H}\mathbf{p}_{\mathbf{x}'}\neq0$,
for all $\mathbf{x}\in\mathcal{C}$ and $\mathbf{x}'\in\Phi\setminus\mathcal{C}$.
This non-orthogonality of the pilot sequences results in a degradation
of the channel estimate quality for the cluster AP channels due to
interference from out-of-cluster APs. One approach to avoid pilot
contamination is to arbitrarily increase $N_{p}$ so as to achieve
(approximate) orthogonality of pilot sequences by application of the
law of large numbers. However, this approach introduces unacceptable
overhead. It is therefore of interest to investigate the system performance
under finite $N_{p}$, a topic that is pursued in the following.

\section{LMMSE Channel Estimation and Data Detection $\mathsf{SNR}$}

This section provides a closed-form expression for the (effective)
data detection $\mathsf{SNR}$, taking into account channel estimation
errors, which serves as a reasonable metric for performance characterization
and design of parameters $N_{a}$ and $N_{p}$. The $\mathsf{SNR}$
expression depends on two important quantities reflecting the NCJT
channel energy and the channel estimation quality, which are investigated
in detail, giving also insights on the optimal $N_{a}$ and $N_{p}$.
The first step towards this investigation is the specification of
the channel estimator, which is discussed next.

\subsection{LMMSE Channel Estimation}

For data detection (decoding), the UE processes the received signal
$y_{d}$ during the data transmission phase under the knowledge of
an estimate $\widehat{\mathbf{1}^{T}\mathbf{h}_{\mathcal{C}}}$ of
the NCJT channel obtained from $\mathbf{y}_{p}$. In this paper, the
standard LMMSE estimator is employed. As the LMMSE estimator is a
Bayesian estimator, the prior information available at the UE side
on the AP channels and pilot sequences must be specified.

Regarding the AP channel information, the standard assumption made
in previous works on Bayesian multipoint channel estimation is that
the AP distances from the typical UE are perfectly known \cite{Kountouris,Bai Heath,cell free vs small cells,Heath channel estimation conf}.
This effectively renders the channel estimation problem equivalent
to the estimation of the Gaussian coefficients $\{c_{\mathbf{x}}\}_{\mathbf{x}\in\mathcal{C}}$
and the specification of the estimator is straightforward. However,
this location information requires a dedicated signaling overhead
to acquire it, which may not be acceptable, at least in the case of
channels with small coherence intervals due to, e.g., mobility of
UEs.

In this paper, a worst-case assumption is considered in terms of prior
information available at the UE side about both the AP channels and
the AP pilot sequences, corresponding to a minimum signaling overhead
required to obtain it. Clearly, analysis under this approach can be
treated as providing a lower performance bound when additional information
is available.

\begin{ass*} 
The typical UE  has no prior information about the small scale fading and location (distance) of any AP in the system, including the APs in the serving cluster. It is only aware of the channel model of (\ref{eq:channel model}) and (\ref{eq:large scale fading law}), the values of $\lambda$, $\alpha$ and $\sigma_w^2$, and the pilot sequences $\{\mathbf{p}_\mathbf{x} \}_{\mathbf{x} \in \mathcal{C}}$ of  the $N_a$ cluster APs.
\end{ass*}

With the prior information at the UE side specified, the LMMSE estimator
can now be obtained.
\begin{prop}
\label{prop: LMMSE estimator formula}The LMMSE estimate of the NCJT
channel equals $\widehat{\mathbf{1}^{T}\mathbf{h}_{\mathcal{C}}}=\mathbf{1}^{T}\hat{\mathbf{h}}_{\mathcal{C}}$,
where
\begin{equation}
\hat{\mathbf{h}}_{\mathcal{C}}\triangleq\left(\frac{N_{a}(\sigma_{w}^{2}+\sigma_{\Phi}^{2}-\sigma_{\mathcal{C}}^{2})}{\sigma_{\mathcal{C}}^{2}}\mathbf{I}_{N_{a}}+\mathbf{P}_{\mathcal{C}}^{H}\mathbf{P}_{\mathcal{C}}\right)^{-1}\mathbf{P}_{\mathcal{C}}^{H}\mathbf{y}_{p},\label{eq:LMMSE estimator}
\end{equation}
is the LMMSE estimate of $\mathbf{h}_{\mathcal{C}}$, $\sigma_{\mathcal{C}}^{2}\triangleq\mathbb{E}(|\mathbf{1}^{T}\mathbf{h}_{\mathcal{C}}|^{2})$
is the (average) NCJT channel energy and
\begin{align}
\sigma_{\Phi}^{2} & \triangleq\mathbb{E}\left(\left|\sum_{\mathbf{x}\in\Phi}h_{\mathbf{x}}\right|^{2}\right)\nonumber \\
 & =\frac{\alpha\lambda\pi r_{0}^{2}}{\alpha-2},\label{eq: total energy}
\end{align}
is the (average) NCJT channel energy when all APs in the system are
included in the cluster ($\mathcal{C}=\Phi$).
\end{prop}
\begin{IEEEproof}
See Appendix \ref{sec:Proof-of-LMMSE estimator formula}.
\end{IEEEproof}
\begin{rem}
By the linearity of the LMMSE estimator \cite{Kay}, estimation of
the NCJT channel $\mathbf{1}^{T}\mathbf{h}_{\mathcal{C}}$ is effectively
equivalent to the estimation of the $N_{a}$ AP channels in $\mathbf{h}_{\mathcal{C}}$.
Note that under the assumed prior information available at the UE,
$\mathbf{h}_{\mathcal{C}}$ is not Gaussian, therefore, its LMMSE
estimator of (\ref{eq:LMMSE estimator}) does not coincide with its
(non-linear) minimum mean square error (MMSE) estimator, which is
much more complicated to compute.

In order to fully specify the LMMSE channel estimator of Proposition
\ref{prop: LMMSE estimator formula}, the NCJT channel energy $\sigma_{\mathcal{C}}^{2}$
must be obtained. This is done after the specification of the data
detection $\mathsf{SNR}$ presented next, as $\sigma_{\mathcal{C}}^{2}$
also affects the $\mathsf{SNR}$ and, therefore, has strong implications
on the data transmission performance.
\end{rem}

\subsection{Data Detection $\mathsf{SNR}$}

For coherent detection purposes, the UE effectively treats $y_{d}$
as equal to \cite{Hassibi}
\begin{equation}
y_{d}=d\widehat{\mathbf{1}^{T}\mathbf{h}_{\mathcal{C}}}+d(\underset{\triangleq e}{\underbrace{\mathbf{1}^{T}\mathbf{h}_{\mathcal{C}}-\widehat{\mathbf{1}^{T}\mathbf{h}_{\mathcal{C}}}})}+w_{d}.\label{eq:I/O system model under chan. est.}
\end{equation}
This is merely a rewriting of (\ref{eq:data I/O equation}), however,
the form of (\ref{eq:I/O system model under chan. est.}) highlights
the assumption made for data detection purposes that data is transmitted
via a channel $\widehat{\mathbf{1}^{T}\mathbf{h}_{\mathcal{C}}}$
(and not $\mathbf{1}^{T}\mathbf{h}_{\mathcal{C}}$, whose value is
unknown) and affected by a noise term $de+w_{d}$ combining the effects
of additive noise and residual channel estimation error. By standard
results from linear estimation theory \cite{Kay}, this noise term
is uncorrelated with the useful signal $d\widehat{\mathbf{1}^{T}\mathbf{h}_{\mathcal{C}}}$.
Therefore, the effective data detection SNR for the signal model of
(\ref{eq:I/O system model under chan. est.}), \emph{conditioned on
the LMMSE channel estimate,} can be defined as
\[
\mathsf{SNR}_{c}\triangleq\frac{\left|\widehat{\mathbf{1}^{T}\mathbf{h}_{\mathcal{C}}}\right|^{2}}{\sigma_{w}^{2}+\sigma_{e}^{2}},
\]
where $\sigma_{e}^{2}\triangleq\mathbb{E}(\left|e\right|^{2})$ is
the mean square error of the LMMSE estimate that is independent of
its actual value \cite{Kay}. By averaging $\mathsf{SNR}_{c}$ over
the channel estimate statistics, an expression for the unconditioned
effective $\mathsf{SNR}$ is immediately obtained as
\begin{align}
\mathsf{SNR} & \triangleq\mathbb{E}\left(\mathsf{SNR}_{c}\right)\nonumber \\
 & =\frac{\sigma_{\mathcal{C}}^{2}-\sigma_{e}^{2}}{\sigma_{w}^{2}+\sigma_{e}^{2}},\label{eq:SNR expression with LMMSE chan. est.}
\end{align}
after using the property $\mathbb{E}(|\chi|^{2})=\mathbb{E}(|\hat{\chi}|^{2})+\mathbb{E}(|\chi-\hat{\chi}|^{2})$,
where $\chi\in\mathbb{C}$ is a random variable and $\hat{\chi}$
its LMMSE estimate. By the ergodicity of the HPPP \cite{Baccelli},
$\mathsf{SNR}$ equals the spatial average $\mathsf{SNR}_{c}$ for
any realization of the AP point process $\Phi$, which can be interpreted
as the value of $\mathsf{SNR}_{c}$ averaged over all UEs in the system
when user-centric NCJT with the same $N_{a}$ is considered for every
UE. Note that $\mathsf{SNR}\leq\sigma_{\mathcal{C}}^{2}/\sigma_{w}^{2}$,
with the upper bound achieved when $\hat{\mathbf{h}}_{\mathcal{C}}=\mathbf{h}_{\mathcal{C}}$
(perfect channel estimation).

Towards further specification of the $\mathsf{SNR}$ formula and obtaining
performance insights, quantities $\sigma_{\mathcal{C}}^{2}$ and $\sigma_{e}^{2}$
appearing in (\ref{eq:SNR expression with LMMSE chan. est.}) are
investigated next.

\subsection{Computation of NJCT Channel Energy\label{subsec:NCJT energy}}

The following result provides an integral-form and an approximate
closed-form expression for $\sigma_{\mathcal{C}}^{2}$, required by
the LMMSE channel estimator as well as for computing the $\mathsf{SNR}$.
\begin{prop}
\label{prop:average cluster energy}The energy of the NCJT channel
equals
\begin{align}
\sigma_{\mathcal{C}}^{2} & ={\normalcolor \frac{2\pi\lambda}{{\color{black}{\color{red}{\color{black}(N_{a}-1)!}}}}}\int_{0}^{\infty}r\ell(r)\Gamma(N_{a},\lambda\pi r^{2})dr\label{eq:cluster energy}\\
 & \approx\sigma_{\Phi}^{2}\left(1-\frac{2}{\alpha}\left(\frac{\lambda\pi r_{0}^{2}}{N_{a}}\right)^{\frac{\alpha}{2}-1}\right),\label{eq: cluster energy approx}
\end{align}
where $\Gamma(a,z)\triangleq\int_{z}^{\infty}t^{a-1}e^{-t}dt$ is
the (lower) incomplete gamma function and the approximation of (\ref{eq: cluster energy approx})
holds for asymptotically large $N_{a}$.
\end{prop}
\begin{IEEEproof}
See Appendix \ref{sec:proof of average cluster energy}.
\end{IEEEproof}
Note that the approximate asymptotic expression of (\ref{eq: cluster energy approx})
is tight in the sense that it reaches the limit $\sigma_{\Phi}^{2}$
as $N_{a}\rightarrow\infty$ ($\mathcal{C}\rightarrow\Phi$). In addition,
(\ref{eq: cluster energy approx}) shows that (a) $\sigma_{\mathcal{C}}^{2}$
is larger for smaller $\alpha$, as expected, due to the smaller propagation
losses, and (b) increasing $N_{a}$ increases $\sigma_{\mathcal{C}}^{2}$,
implying that, under perfect CSI, NCJT with $N_{a}>1$ provides an
$\mathsf{SNR}$ gain compared to conventional, non-cooperative transmission
($N_{a}=1$).

\begin{figure}[t]
\noindent \begin{centering}
\includegraphics[width=1\columnwidth]{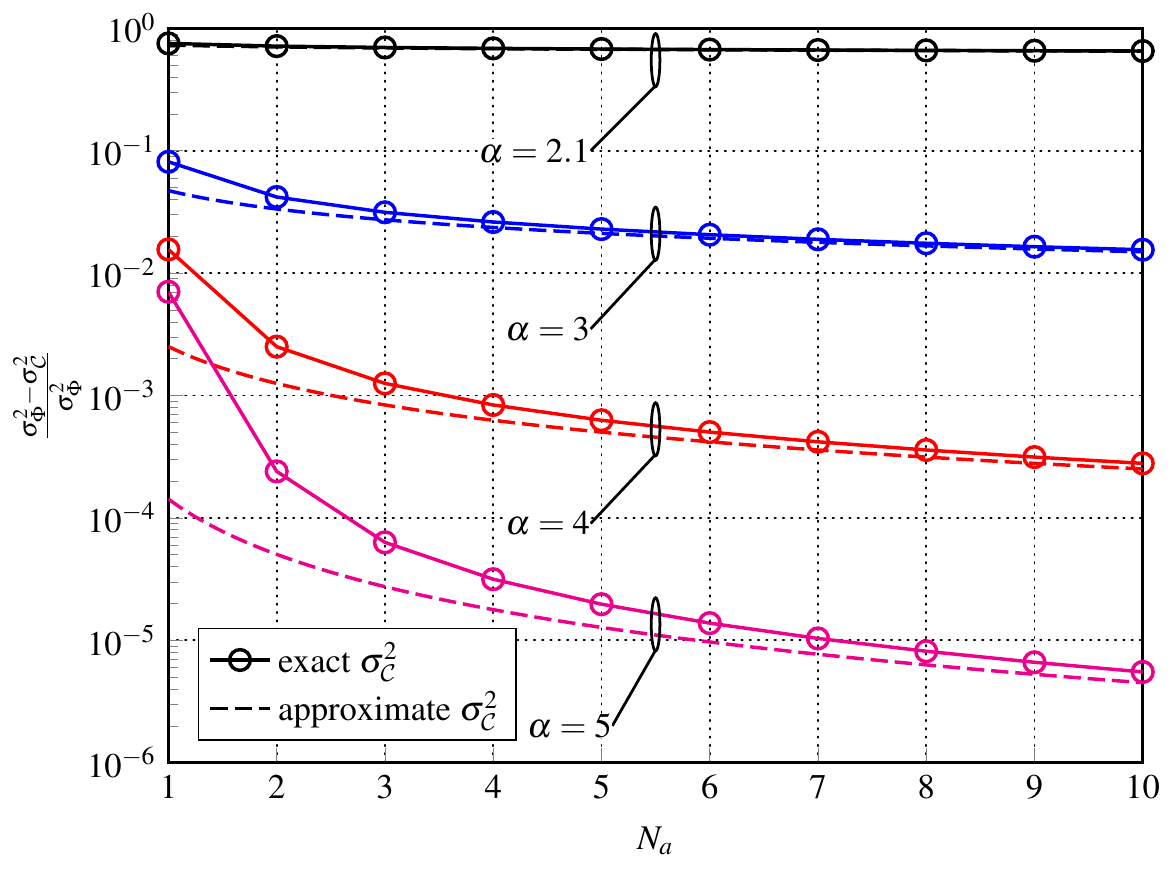}
\par\end{centering}
\caption{\label{fig: cluster energy} Normalized NCJT channel energy difference
with an infinite and finite size AP cluster. The curves for the exact
and approximate expressions are almost exactly the same for $\alpha=2.1$.}
\end{figure}

Figure \ref{fig: cluster energy} shows $(\sigma_{\Phi}^{2}-\sigma_{\mathcal{C}}^{2})/\sigma_{\Phi}^{2}$
as a function of $N_{a}$ for various values of $\alpha$, using both
the exact and approximate expressions for $\sigma_{\mathcal{C}}^{2}$.
This quantity measures how smaller the NCJT channel energy of a finite
AP cluster is from the case with all APs included in the cluster.
For this example, as well as all the numerical examples in this paper,
the reference distance was set to $r_{0}=0.08/(2\sqrt{\lambda})$,
i.e., $8/100$ of the average distance from the closest AP. This corresponds
to a probability $1-e^{-\lambda\pi r_{0}^{2}}\approx5\times10^{-3}$
of an AP in the system having a distance from the typical UE smaller
than $r_{0}$. Note that (\ref{eq: cluster energy approx}) suggests
that $\sigma_{\mathcal{C}}^{2}$ becomes conveniently independent
of $\lambda$ when $r_{0}=\text{const.}/\sqrt{\lambda}$, which can
actually be shown to be the case also for the exact expression of
(\ref{eq:cluster energy}).

It can be seen from Fig. \ref{fig: cluster energy} that the asymptotic
expression of (\ref{eq: cluster energy approx}) is a very good indicator
of the actual $\sigma_{\mathcal{C}}^{2}$, even for moderate values
of $N_{a}$. As predicted by (\ref{eq: cluster energy approx}), increasing
$N_{a}$ increases $\sigma_{\mathcal{C}}^{2}$ towards $\sigma_{\Phi}^{2}$,
however, this increase is only marginal. This is because, in order
to capture most of the maximum possible NCJT channel energy, a value
of $N_{a}=1$ is sufficient when $\alpha$ is large, as the average
received energy from even the second closest AP is much smaller than
that of the closest AP, whereas a very large $N_{a}$ is required
when $\alpha$ is small, since the signals of even distant APs are
strongly received. This observation suggests that the conventional,
non-cooperative, cellular network operation ($N_{a}=1$) is practically
sufficient under perfect CSI when the $\mathsf{SNR}$ is the figure
of merit for the system. However, as will be shown in the following,
$N_{a}>1$ can provide significant $\mathsf{SNR}$ gains in the presence
of channel estimation errors.

\subsection{\label{subsec:Computation of error variance}Computation of LMMSE
Channel Estimation Error Variance}

\begin{figure}[t]
\noindent \centering{}\includegraphics[width=1\columnwidth]{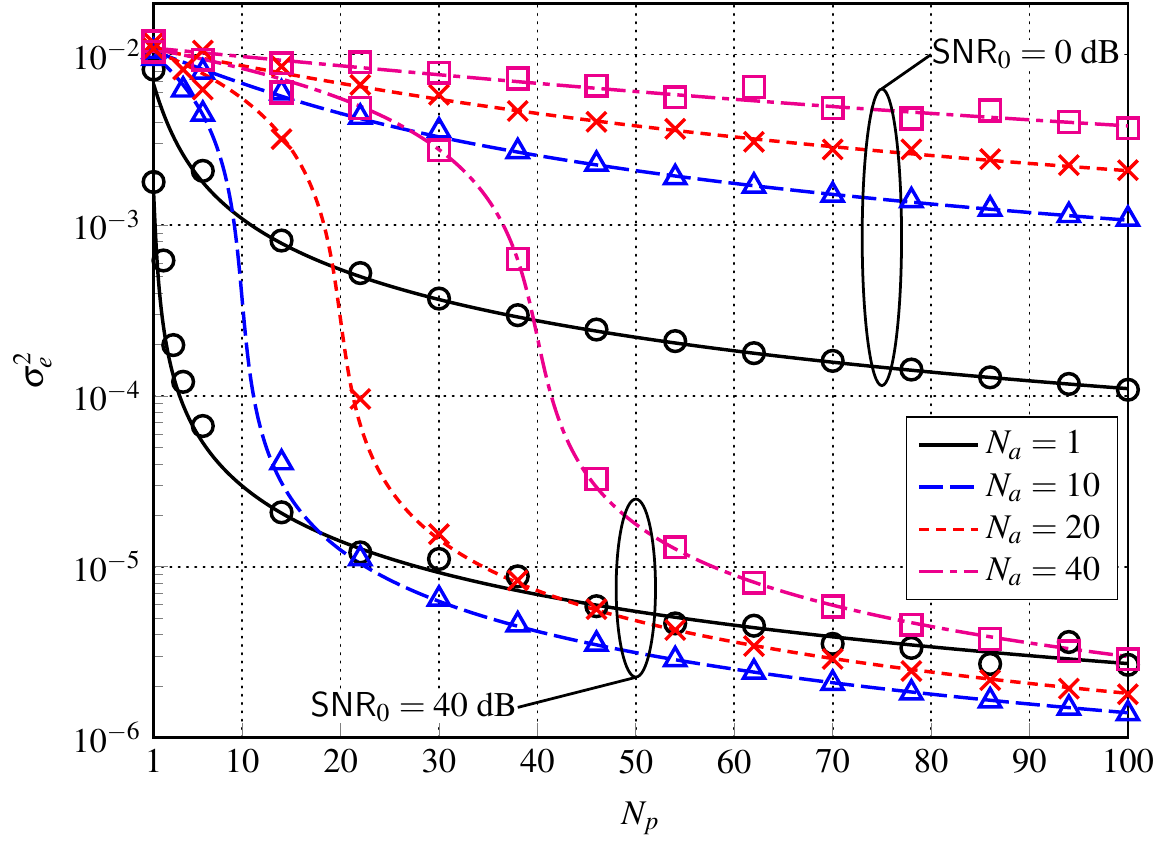}\caption{\label{fig:var_e vs N_p} Channel estimation error variance as a function
of the pilot sequences length. Markers denote Monte Carlo simulation
results and the approximate expression of (\ref{eq:error variance})
is plotted as a continuous function of $N_{p}$ ($\alpha=3.67$).}
\end{figure}

The following result provides a closed form asymptotic approximation
for the channel estimation error variance $\sigma_{e}^{2}$ required
for the  $\mathsf{SNR}$ computation.
\begin{prop}
\label{prop:channel estimation variance}For $N_{p},N_{a}\rightarrow\infty$
with the ratio $N_{a}/N_{p}$ constant, the channel estimation mean
square error approximately equals
\begin{align}
\sigma_{e}^{2} & =\frac{N_{a}(\sigma_{w}^{2}+\sigma_{\Phi}^{2}-\sigma_{\mathcal{C}}^{2})}{N_{p}}f_{N_{a}/N_{p}}\left(\frac{\sigma_{w}^{2}+\sigma_{\Phi}^{2}-\sigma_{\mathcal{C}}^{2}}{\sigma_{\mathcal{C}}^{2}}\right),\label{eq:error variance}
\end{align}
where $f_{a}(b)\triangleq\frac{a-1+\sqrt{(1+a-ab)^{2}-4a^{2}b}}{2a^{2}b}-\frac{1}{2a}$.
\end{prop}
\begin{IEEEproof}
See Appendix \ref{sec: proof of channel estimation variance}.
\end{IEEEproof}
As noted in the proof of this result, an exact, non-asymptotic expression
for $\sigma_{e}^{2}$ is available, which, however, is a function
of the random pilot sequences matrix $\mathbf{P}_{\mathcal{C}}$.
The asymptotic expression of (\ref{eq:error variance}) is much more
convenient as it applies to every realization of $\mathbf{P}_{\mathcal{C}}$
and is also very accurate even for small values of $N_{p}$ and $N_{a}$.
This is verified in Fig. \ref{fig:var_e vs N_p} where $\sigma_{e}^{2}$
is plotted as a function of $N_{p}$, evaluated both using Monte Carlo
simulation and the expression of (\ref{eq:error variance}). The simulation
results are obtained by averaging the channel estimation square error
$|e|^{2}$ over independent realizations of $\Phi$, $\{h_{\mathbf{x}}\}_{\mathbf{x}\in\Phi}$,
and $\{\mathbf{p}_{\mathbf{x}}\}_{\mathbf{x}\in\Phi}$. As (\ref{eq:error variance})
is independent of $\lambda$, a value of $\lambda=1$ was arbitrarily
chosen for generating the AP distribution realizations. Various values
of $N_{a}$ were considered and $\sigma_{w}^{2}$ was set such that
$\mathsf{SNR}_{0}=\{0,40\}$ dB, where $\mathsf{SNR}_{0}\triangleq\mathbb{E}(|h_{\mathbf{x}_{1}}|^{2})/\sigma_{w}^{2}$
is the $\mathsf{SNR}$ achieved with $N_{a}=1$ and perfect CSI ($h_{\mathbf{x}_{1}}$
is the channel of the closest AP). Note that $\mathbb{E}(|h_{\mathbf{x}_{1}}|^{2})$
can be computed from Proposition \ref{prop:average cluster energy}
as $\sigma_{\mathcal{C}}^{2}$ with $N_{a}=1$. These two $\mathsf{SNR}_{0}$
values were chosen as representative of operational conditions were
the effect of additive noise is dominant (power limited regime) and
the effect of interference is dominant (pilot contamination regime)
in the channel estimation procedure, respectively. The path loss factor
was set to $\alpha=3.67$, a value commonly considered by cellular
standards \cite{Ahmadi LTE book}, with similar results observed for
other values of $\alpha$.

As expected, the channel estimation mean square error decreases with
increasing $N_{p}$ in all cases. In the power limited regime ($\mathsf{SNR}_{0}=0$
dB), the NCJT channel estimate degrades with increasing $N_{a}$,
irrespective of the pilot sequence length $N_{p}$. This is because
attempting to estimate more parameters (in this case, AP channels)
in the presence of additive noise can only degrade performance \cite{Kay},
even more so since the channels corresponding to distant APs have
(very) small energy. However, when the channel estimation is mostly
affected by pilot contamination ($\mathsf{SNR}_{0}=40$ dB), the optimal
$N_{a}$ has to balance two conflicting requirements: (a) estimate
only a few channels corresponding to strongly received AP signals,
and (b) consider a large cluster size that results in a statistically
small out-of-cluster interference. As can be seen in Fig. \ref{fig:var_e vs N_p},
the optimal $N_{a}$ for $\mathsf{SNR}_{0}=40$ dB depends on $N_{p}$
and becomes greater than $1$ for large $N_{p}$. Given that a greater
$N_{a}$ also corresponds to a greater NJCT channel energy $\sigma_{\mathcal{C}}^{2}$,
it directly follows from (\ref{eq:SNR expression with LMMSE chan. est.})
that $N_{a}>1$ is optimal w.r.t. $\mathsf{SNR}$ for sufficiently
large $N_{p}$ in the pilot contamination regime.

\section{On the Optimal Pilot Sequence Length and AP Cluster Size}

Propositions \ref{prop:average cluster energy} and \ref{prop:channel estimation variance}
allow for a simple computation of the $\mathsf{SNR}$ without the
need to resort to computationally intensive Monte Carlo simulations.
This, in turn, allows for efficient (numerical) optimization of the
two fundamental system design parameters $N_{p}$ and $N_{a}$. However,
some interesting analytical insights under specific operational conditions
can be obtained.

\subsection{\label{subsec:Minimum-Required-Pilot-Length}Minimum Required Pilot
Sequence Length}

Consideration of the $\mathsf{SNR}$ as a metric for identifying the
optimal $N_{p}$ leads to impractical designs since, clearly, the
maximum $\mathsf{SNR}$ performance corresponding to perfect CSI is
achieved with $N_{p}\rightarrow\infty$. Towards a practical design
rule, it is reasonable to look for the minimum $N_{p}$ required to
achieve an $\mathsf{SNR}$ that is a given fraction of the perfect
CSI $\mathsf{SNR}$. The following result provides a simple expression
for this value.
\begin{prop}
\label{prop:Minimum pilot length}For fixed cluster size $N_{a}$
and $0<\gamma<1$ such that it holds
\begin{equation}
\gamma\gg\frac{1}{N_{a}+1},\label{eq:N_p condition}
\end{equation}
the (minimum) pilot sequence length required to achieve $\mathsf{SNR}\geq\gamma\sigma_{\mathcal{C}}^{2}/\sigma_{w}^{2}$
approximately equals
\begin{align}
N_{p}^{*} & =\frac{\gamma N_{a}(\sigma_{\Phi}^{2}-\sigma_{\mathcal{C}}^{2}+\sigma_{w}^{2})(\sigma_{\mathcal{C}}^{2}+\sigma_{w}^{2})}{(1-\gamma)\sigma_{\mathcal{C}}^{2}\sigma_{w}^{2}}\label{eq:N_p formula}\\
 & >\frac{\gamma N_{a}}{1-\gamma}\label{eq:N_p formula bound}
\end{align}
\end{prop}
\begin{IEEEproof}
See Appendix \ref{sec:Proof of minimum pilot length proposition}.
\end{IEEEproof}
Note that condition (\ref{eq:N_p condition}) holds when $\gamma$
is selected sufficiently close to $1$ and/or $N_{a}$ is sufficiently
small, which is typically the case in system design towards minimizing
degradation due to channel estimation error and minimizing the physical
resources (APs) dedicated to each UE.

The simple bound of (\ref{eq:N_p formula bound}) suggests that $N_{p}^{*}$
must be greater than a value that is proportional to $N_{a}$, as
expected, however, with a proportionality constant $\gamma/(1-\gamma)$
that becomes exponentially large as $\gamma\rightarrow1$, i.e., achieving
performance extremely close to the perfect CSI case requires excessively
large overhead irrespective of the operational conditions. In addition,
the closed-form expression of (\ref{eq:N_p formula}) reveals that
$N_{p}^{*}$ is a convex function of the noise variance $\sigma_{w}^{2}$
with its extremal behavior following by simple inspection.
\begin{cor}
In the power limited ($\sigma_{w}^{2}\rightarrow\infty$) and pilot
contamination ($\sigma_{w}^{2}\rightarrow0$) operational regimes,
$N_{p}^{*}\rightarrow\infty$.
\end{cor}
The unbounded pilot overhead as $\sigma_{w}^{2}\rightarrow\infty$
and $\sigma_{w}^{2}\rightarrow0$ is required in order to average
the noise effects in the power limited regime and orthogonalize the
AP pilot sequences in the pilot contamination regime. Noting that
the perfect CSI $\mathsf{SNR}$ in the power limited regime is fundamentally
small, it follows that operation in this regime is highly undesirable
both from a pilot overhead and achieved $\mathsf{SNR}$ perspective.
This is essentially a system level manifestation of the well known
fact that pilot-aided coherent processing in point-to-point communications
is highly suboptimal in the power limited regime \cite{Hassibi}.
Although the pilot overhead requirements are also very large in the
pilot contamination regime, it is noted that, since the perfect CSI
$\mathsf{SNR}$ is extremely large there (approaching infinity), there
is no practical need to consider a large $\gamma$ as in (\ref{eq:N_p condition}),
and a moderate value of $N_{p}$ would be sufficient.

\subsection{Optimal Cluster Size with Arbitrarily Large Training Overhead}

Another important design question is the identification of the optimal
cluster size $N_{a}$ given that a pilot sequence length $N_{p}$
can be afforded. As was discussed in Sec. \ref{subsec:Computation of error variance},
a cluster with $N_{a}>1$ APs is $\mathsf{SNR}$-optimal in the pilot
contamination regime with sufficiently large $N_{p}$. Intuitively,
this is because a larger $N_{p}$ allows for the accurate estimation
of multiple AP channels, which, in turn, also results in a larger
cluster and a reduced out-of-cluster interference. One may then wonder
whereas this implies that $N_{a}$ becomes arbitrarily large when
an arbitrarily large $N_{p}$ can be afforded. The following result
shows that this is the case only for a sufficiently large path loss
factor.
\begin{prop}
\label{prop: optimal cluster size}When an arbitrarily large (but
finite) pilot sequence length $N_{p}$ can be afforded, the AP cluster
size $N_{a}$ that maximizes $\mathsf{SNR}$ with $\sigma_{w}^{2}=0$
(pilot contamination regime) is bounded for $\alpha\leq4$ and arbitrarily
large for $\alpha>4$.
\end{prop}
\begin{IEEEproof}
See Appendix \ref{sec:Proof of optimal cluster size}.
\end{IEEEproof}
This result can be explained by noting that, for small $\alpha$,
the total out-of-cluster interference will always be very strong,
irrespective of the cluster size, rendering the estimation of too
many AP channels, most of them of small energy, highly inaccurate.
Interestingly, the path loss factor $\alpha=4$ is identified as a
phase transition threshold from a bounded to an unbounded value for
the optimal $N_{a}$.
\begin{figure}[t]
\noindent \begin{centering}
\includegraphics[width=1\columnwidth]{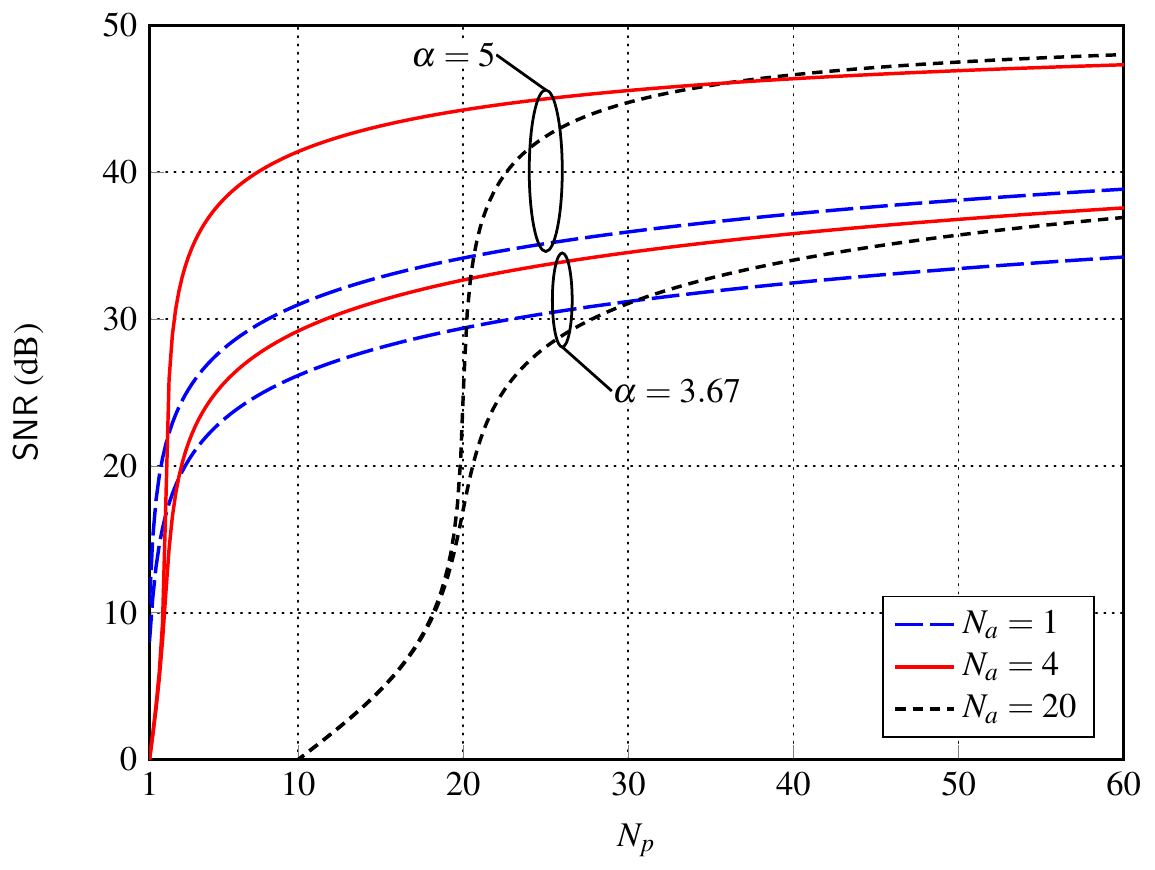}
\par\end{centering}
\caption{\label{fig: SNR vs Np}$\mathsf{SNR}$ as a function of pilot sequence
length $N_{p}$ ($\mathsf{SNR}_{0}=50$ dB).}
\end{figure}

\begin{rem}
It follows from the proof of Proposition \ref{prop: optimal cluster size}
that, when $\sigma_{w}^{2}=0$, the $\mathsf{SNR}$ admits the simple
expression 
\begin{equation}
\mathsf{SNR}=\frac{N_{p}\sigma_{\mathcal{C}}^{2}}{N_{a}(\sigma_{\Phi}^{2}-\sigma_{\mathcal{C}}^{2})}\left(1-\mathcal{O}\left(\frac{N_{a}}{N_{p}}\right)\right),\frac{N_{p}}{N_{a}}\rightarrow\infty,\label{eq:SNR expression pilot contamination regime}
\end{equation}
that is valid for all $\alpha$, clearly showing that, for any fixed
cluster size $N_{a}$, the $\mathsf{SNR}$ increases proportionally
to $N_{p}$.
\end{rem}

\section{Numerical Examples}

This section considers a number of representative numerical examples
providing insights on the $\mathsf{SNR}$ properties and the optimization
of $N_{p}$ and/or $N_{a}$. Unless stated otherwise, the $\mathsf{SNR}$
is computed using the analytical expression of (\ref{eq:SNR expression with LMMSE chan. est.})
with the exact $\sigma_{\mathcal{C}}^{2}$ expression of (\ref{eq:cluster energy})
and with $\sigma_{e}^{2}$ as in (\ref{eq:error variance}). As in
Secs. \ref{subsec:NCJT energy} and \ref{subsec:Computation of error variance},
the reference distance was set to $r_{0}=0.08/(2\sqrt{\lambda})$,
which renders the results independent of $\lambda$ (a value of $\lambda=1$
was arbitrarily used for the Monte Carlo simulations).

\emph{1) $\mathsf{SNR}$ dependence on $N_{p}$ and $N_{a}$}: Figure
\ref{fig: SNR vs Np} shows the $\mathsf{SNR}$ as a function of $N_{p}$
for a few representative values of $N_{a}$ and $\alpha$. The noise
variance was set such that $\mathsf{SNR}_{0}=50$ dB (pilot contamination
regime). It can be seen that in all cases, increasing $N_{p}$ increases
$\mathsf{SNR}$ due to the corresponding improvement of the NCJT channel
estimate. For the parameters considered, a trivial cluster size with
$N_{a}=1$ is the best only for very small values of $N_{p}$. Larger
values of $N_{p}$ favor greater $N_{a}$, in line with the observations
made in Sec. \ref{subsec:Computation of error variance} regarding
channel estimation error. In addition, the $\mathsf{SNR}$ gain achieved
by using $N_{a}>1$ instead of $N_{a}=1$ is more pronounced for larger
$\alpha$. For the parameters considered in this example, a gain of
about $10$ dB and $3.5$ dB is achieved when $\alpha=5$ and $3.67$,
respectively, for all $N_{p}>10$.

\begin{figure}[t]
\noindent \centering{}\includegraphics[width=1\columnwidth]{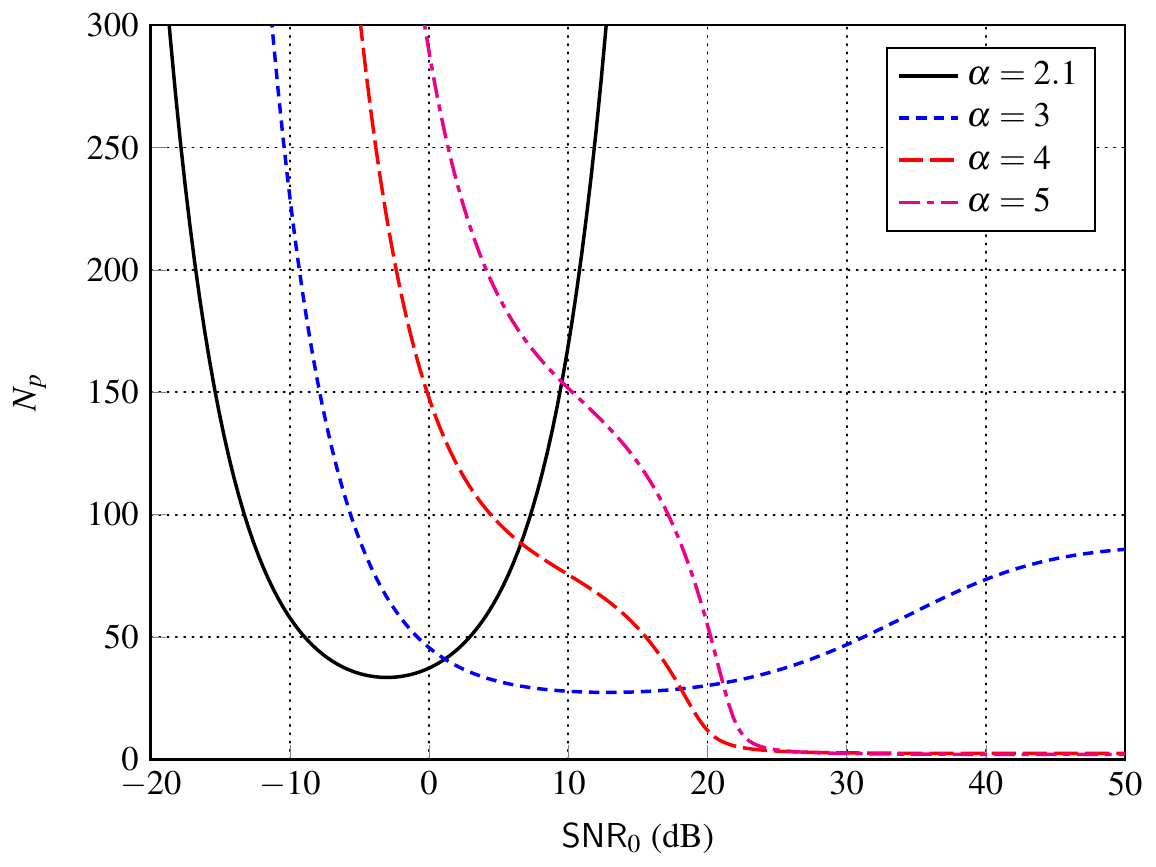}\caption{\label{fig: NCJT region} Minimum required $N_{p}$ for achieving
a greater $\mathsf{SNR}$ with $N_{a}=2$ than with $N_{a}=1$.}
\end{figure}

\begin{figure*}[t]
\noindent \centering{}\subfloat[$\alpha=3.67$]{\noindent \begin{centering}
\includegraphics[width=1\columnwidth]{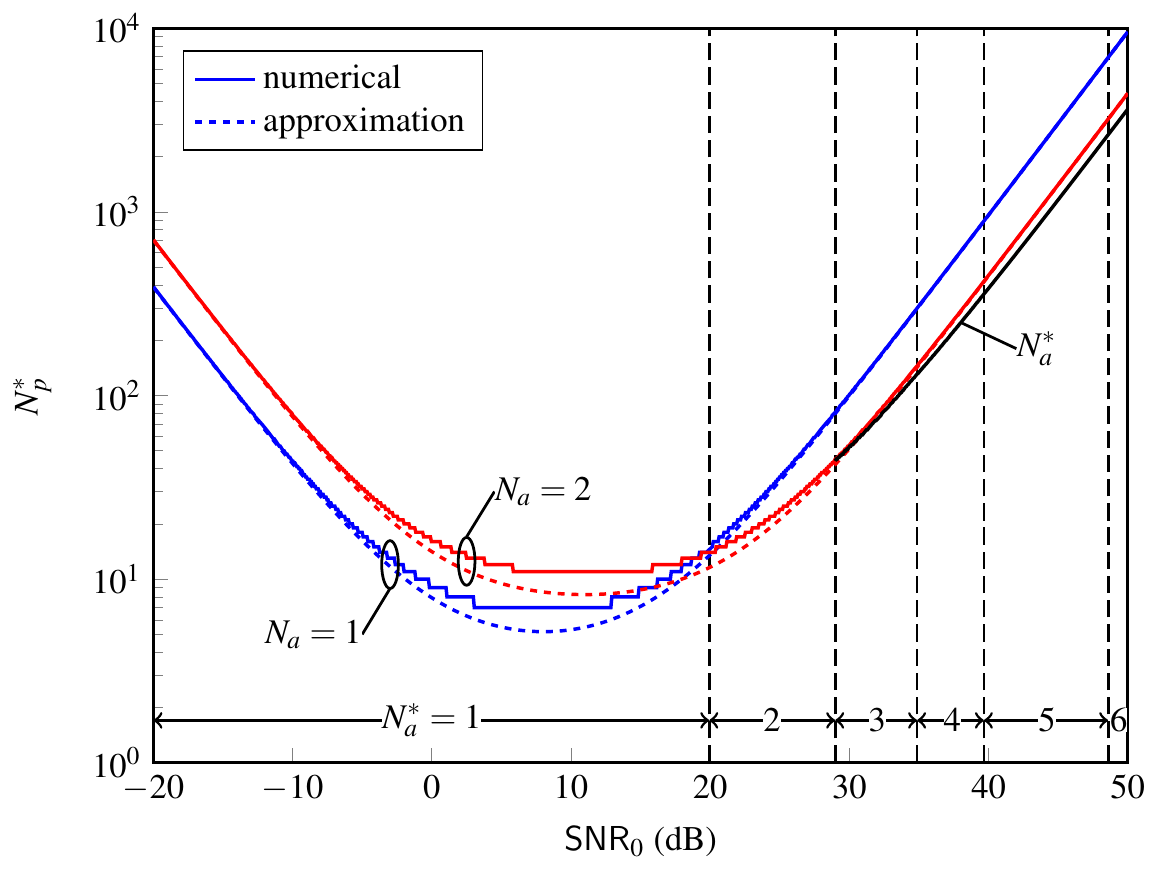}
\par\end{centering}
}\subfloat[$\alpha=5$]{\noindent \centering{}\includegraphics[width=1\columnwidth]{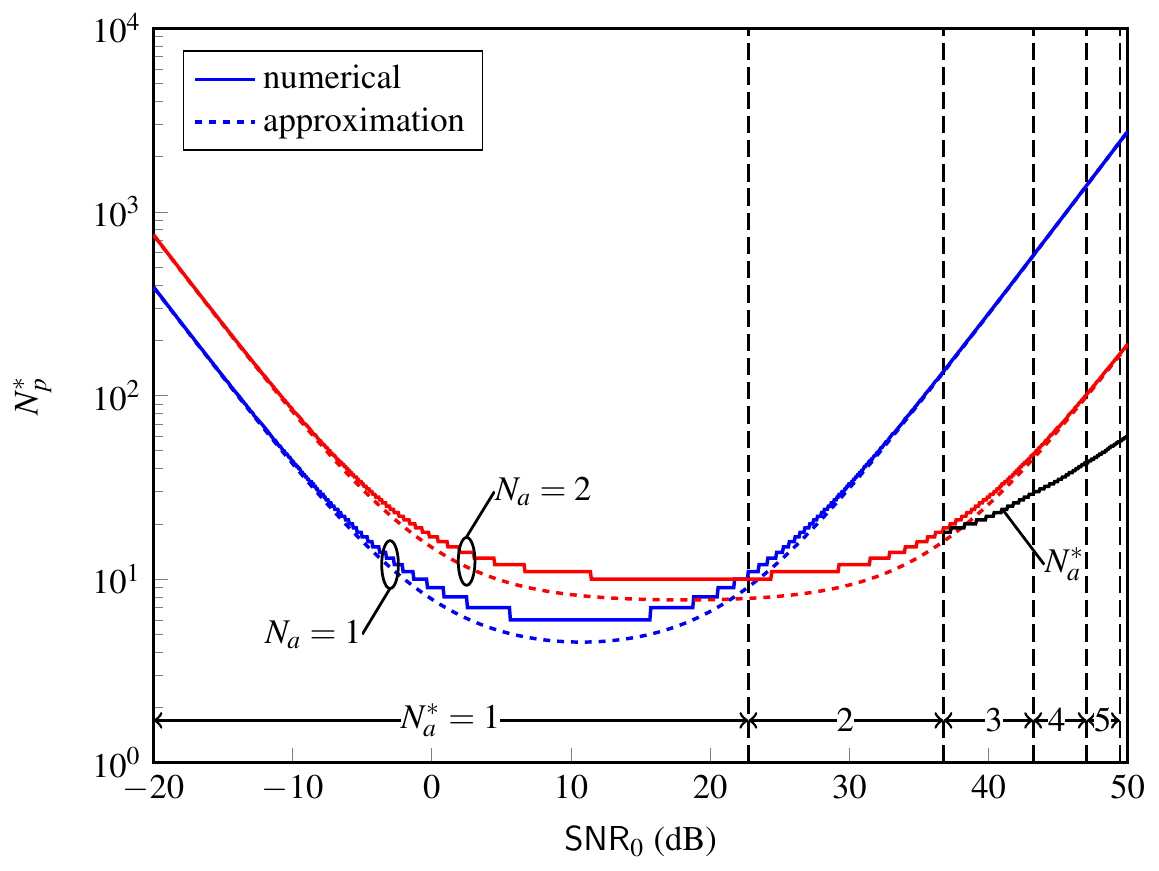}}\caption{\label{fig:minimum Np over SNR0} Minimum $N_{p}$ required to achieve
an $\mathsf{SNR}$ that is $1$ dB less than $\mathsf{SNR}_{0}$.
The cases of $N_{a}=1$, $2$, as well as optimal cluster size $N_{a}^{*}$
are depicted.}
\end{figure*}

\emph{2) Optimality of multipoint transmissions}: As observed previously,
a sufficiently large $N_{p}$ will render a multipoint cluster size
($N_{a}>1$) $\mathsf{SNR}$-optimal in the pilot contamination regime.
In order to gain insights on what values of $N_{p}$ promote multipoint
transmissions, Fig. \ref{fig: NCJT region} plots the value of $N_{p}$
above which $N_{a}=2$ provides a larger $\mathsf{SNR}$ than $N_{a}=1$.
A wide range of $\mathsf{SNR_{0}}$ values is considered covering
the pilot contamination as well as the power limited regime. For visualization
purposes, the plotted $N_{p}$ values were obtained as the real number
for which the $\mathsf{SNR}$ for $N_{a}=1$ and $N_{a}=2$ is the
same. It can be seen that both the operational regime as well as the
path loss factor $\alpha$ critically affect the optimality of multipoint
transmissions and required $N_{p}$. In the power contamination regime
(large $\mathsf{SNR}_{0}$), the minimum $N_{p}$ for which $N_{a}=2$
becomes optimal ranges from extremely large to as small as $1$ with
increasing $\alpha$. Essentially, multipoint transmission is always
optimal in this regime for values of $\alpha$ greater than $4$.
On the contrary, in the power limited regime (small $\mathsf{SNR}_{0}$),
smaller values of $N_{p}$ for which $N_{a}=2$ is optimal are favored
by smaller $\alpha$, although the corresponding $N_{p}$ never falls
below $40$ and becomes arbitrarily large as $\mathsf{SNR_{0}}$ further
decreases. This suggests that $N_{a}=1$ is optimal in the pilot limited
regime for all $\alpha$.

\emph{3) Dependence of minimum $N_{p}$ on $\sigma_{w}^{2}$ and $N_{a}$}:
As a design problem example, consider the identification of the (minimum)
$N_{p}$, $N_{p}^{*}$, required to achieve an $\mathsf{SNR}$ that
is $1$ dB less than $\mathsf{SNR_{0}}$, the $\mathsf{SNR}$ achieved
with $N_{a}=1$ and perfect CSI. Note that this design entails solving
the inequality $\mathsf{SNR}\geq10^{-1/10}\mathsf{SNR}_{0}$ w.r.t.
$N_{p}$, which can be efficiently done numerically using the analytical
$\mathsf{SNR}$ expression. Figure \ref{fig:minimum Np over SNR0}
shows the solution as a function of $\mathsf{SNR}_{0}$ for two cases
of $\alpha$ and for $N_{a}=1,2,$ as well as $N_{a}=N_{a}^{*}$,
where $N_{a}^{*}$ is the value of $N_{a}$, depending on $\mathsf{SNR_{0}}$,
that leads to a minimum $N_{p}^{*}$. The value of $N_{a}^{*}$ is
also indicated over the $\mathsf{SNR}_{0}$ range. In addition, the
approximate $N_{p}^{*}$ expression of (\ref{eq:N_p formula}) with
$\gamma=10^{-1/10}$ and $\gamma=10^{-1/10}\sigma_{2}^{2}/\sigma_{1}^{2}$,
for $N_{a}=1$ and $2$, respectively, is also shown, ($\sigma_{i}^{2},i=1,2$,
representing the NCJT channel energy with $N_{a}=i$).

It can be seen that the approximate closed-form expression for $N_{p}^{*}$
is a very good match to the actual $N_{p}^{*}$. In accordance to
the discussion in Sec. \ref{subsec:Minimum-Required-Pilot-Length},
$N_{p}^{*}$ is a convex function of $\mathsf{SNR}_{0}$, increasing
arbitrarily as either $\mathsf{SNR}_{0}$ decreases (power limited
regime) or increases (pilot contamination regime). It can also be
seen that for all $\mathsf{SNR}_{0}$ less than about $20$ dB, $N_{a}^{*}=1$
for both values of $\alpha$. In contrast, for increasing $\mathsf{SNR}_{0}$,
clusters of $N_{a}>1$ APs are preferable with $N_{a}^{*}$ also increasing
as $\mathsf{SNR}_{0}$ increases. Similar to the $\mathsf{SNR}$ behavior
discussed in the first example, the gain in pilot overhead reduction
by using $N_{a}>1$ instead of $N_{a}=1$ is more prominent for greater
$\alpha$. Even with a cluster of only $N_{a}=2$ APs, over an order
of magnitude smaller $N_{p}$ is required when $\mathsf{SNR}_{0}>40$
dB and $\alpha=5$.
\begin{figure}[t]
\noindent \centering{}\includegraphics[width=1\columnwidth]{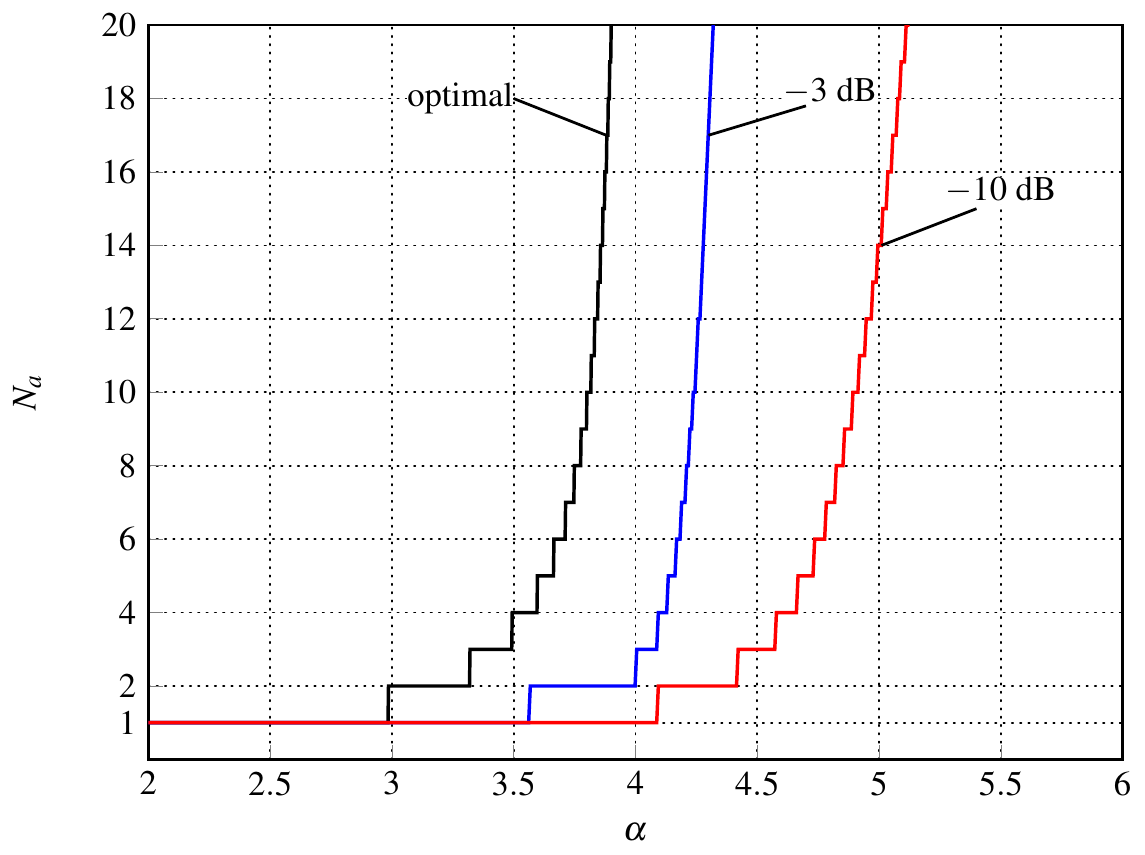}\caption{\label{fig:Optimal Na vs alpha} $\mathsf{SNR}$-optimal and suboptimal
$N_{a}$ (resulting in a $3$ dB and $10$ dB $\mathsf{SNR}$ loss)
in the pilot contamination regime with arbitrarily large $N_{p}$
.}
\end{figure}

\emph{4) $\mathsf{SNR}$-optimal $N_{a}$ for $N_{p}\gg1$ and $\sigma_{w}^{2}=0$}:
In the pilot contamination regime ($\sigma_{w}^{2}=0$), it directly
follows from (\ref{eq:SNR expression pilot contamination regime}),
that, for any $N_{p}\gg1$, the $\mathsf{SNR}$-optimal cluster size
$N_{a}^{*}$ is the one that maximizes the term $\frac{\sigma_{\mathcal{C}}^{2}}{N_{a}(\sigma_{\Phi}^{2}-\sigma_{\mathcal{C}}^{2})}$
(note that $\sigma_{\mathcal{C}}^{2}$ depends on $N_{a}$). The result
of this maximization is shown in Fig. \ref{fig:Optimal Na vs alpha},
where the optimal $N_{a}$ is depicted as a function of the path loss
factor $\alpha$. The phase transition threshold of $\alpha=4$ identified
in Proposition \ref{prop: optimal cluster size} is also verified
numerically, with values of $\alpha>4$ corresponding to an arbitrarily
large optimal $N_{a}$. The optimal $N_{a}$ becomes finite for $\alpha\leq4,$
although it is still very large for values of $\alpha$ close to $4$,
and decreases to $N_{a}=1$ as $\alpha$ becomes closer to $2$. For
the practical value of $\alpha=3.67$, $N_{a}^{*}=8$.

Towards reducing the physical resources dedicated to a UE, it is of
interest to investigate a potential reduction of $N_{a}$ when a suboptimal
$\mathsf{SNR}$ performance is acceptable. This $N_{a}$ can be obtained
as the minimum $N_{a}$ satisfying
\begin{equation}
\frac{\sigma_{\mathcal{C}}^{2}}{N_{a}(\sigma_{\Phi}^{2}-\sigma_{\mathcal{C}}^{2})}\geq\gamma\max_{N_{a}}\frac{\sigma_{\mathcal{C}}^{2}}{N_{a}(\sigma_{\Phi}^{2}-\sigma_{\mathcal{C}}^{2})}\label{eq: suboptimal SNR Na selection}
\end{equation}
for some pre-selected $0<\gamma<1$. The result is shown in Fig. \ref{fig:Optimal Na vs alpha}
for $\gamma=10^{-3/10}$ and $\gamma=10^{-10/10}$, corresponding
to a $3$ dB and $10$ dB $\mathsf{SNR}$ degradation from the optimal.
For values of $\alpha$ corresponding to an extremely large $\mathsf{SNR}$-optimal
$N_{a}$, the right-hand side of (\ref{eq: suboptimal SNR Na selection})
was approximated as equal to the value of $\frac{\gamma\sigma_{\mathcal{C}}^{2}}{N_{a}(\sigma_{\Phi}^{2}-\sigma_{\mathcal{C}}^{2})}$
with $N_{a}=1000$. As expected, allowing for suboptimal $\mathsf{SNR}$
performance results in a decrease of $N_{a}^{*}$, even for $\alpha>4$.
This decrease is greater the larger the allowed $\mathsf{SNR}$ degradation
is.

\begin{figure}[t]
\noindent \centering{}\includegraphics[width=1\columnwidth]{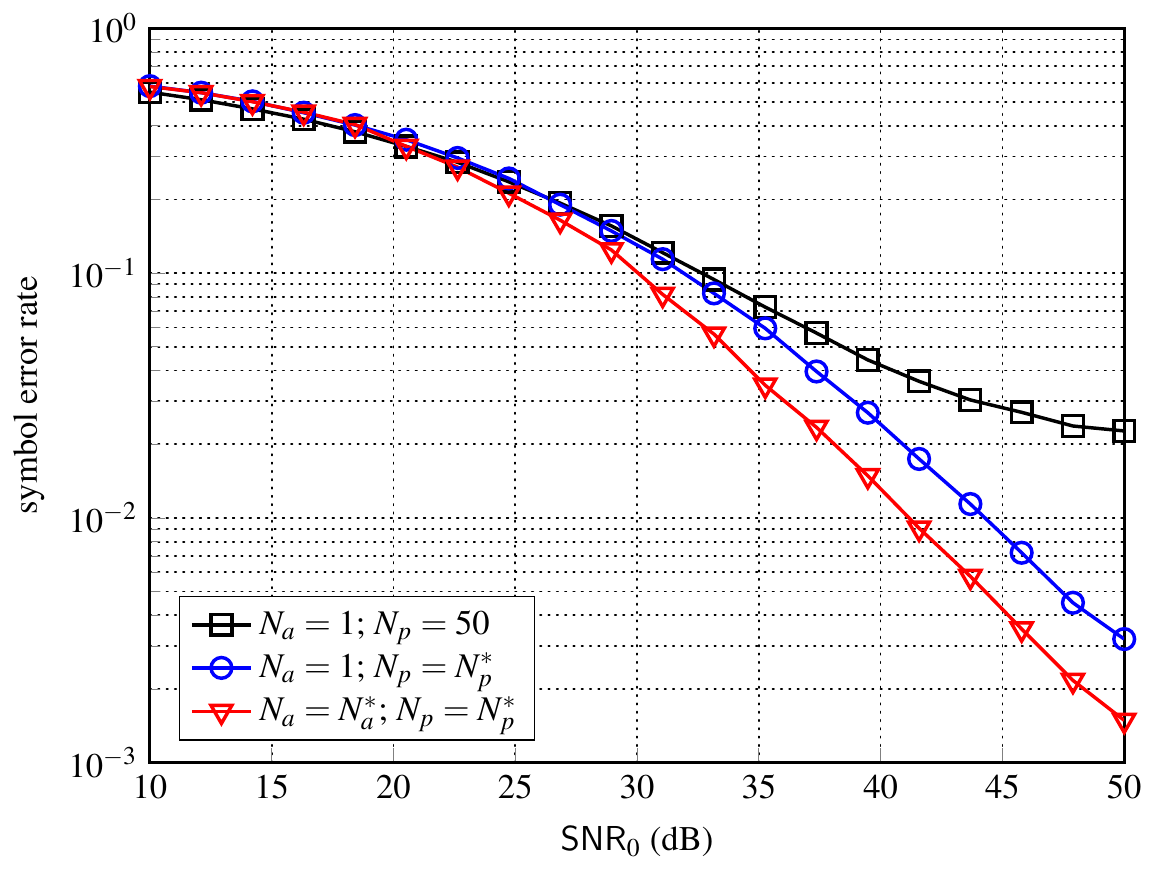}\caption{Symbol error rate with uncoded QPSK modulation ($\alpha=3.67$).\label{fig:SER performance}}
\end{figure}

\emph{5) Symbol error rate performance}: To illustrate the applicability
of the $\mathsf{SNR}$-based design also w.r.t. other communication-theoretic
metrics, Fig. \ref{fig:SER performance} shows the symbol error rate
(SER) experienced at the typical UE when uncoded quadrature phase
shift keying (QPSK) modulation is considered. The receiver first equalizes
the received signal using the LMMSE channel estimate of (\ref{eq:LMMSE estimator}),
i.e., it computes $y_{d}/\widehat{\mathbf{1}^{T}\mathbf{h}_{\mathcal{C}}}$,
and declares the closest in Euclidean distance QPSK symbol as the
transmitted one. Note that this approach is not optimal for minimization
of the SER, however, it is attractive due to its simplicity. The SER
is obtained by Monte Carlo simulation, over a range of moderate to
large $\mathsf{SNR}_{0}$ values corresponding to operation in the
pilot contamination regime. The following schemes were considered
regarding cluster size $N_{a}$ and pilot sequence length $N_{p}$:
(a) $N_{a}=1$ and $N_{p}=50$, (b) $N_{a}=1$ and $N_{p}$ equal
to the value $N_{p}^{*}$ resulting in an $\mathsf{SNR}$ loss of
$1$ dB compared to the perfect CSI $\mathsf{SNR}$, and (c) for the
same $N_{p}$ as case (b), the cluster size $N_{a}^{*}$ that maximizes
$\mathsf{SNR}$ was selected. The path loss factor was set to $\alpha=3.67$.

It can be seen that the first scheme results in a significant error
floor due to insufficient training for compensating the pilot contamination
effect in the high $\mathsf{SNR_{0}}$ regime. In contrast, adapting
$N_{p}$ to $\mathsf{SNR}_{0}$ as per the second scheme eliminates
this effect. Of course, a (very) large fixed value for $N_{p}$ could
be considered that would eliminate the error floor as well, however,
it would have to be chosen by a trial-and-error procedure and would
result in unnecessary signaling overhead under small and moderate
$\mathsf{SNR}_{0}$ conditions. When a cluster of more than one AP
is allowed for the same $N_{p}$, SER improves in the high $\mathsf{SNR}_{0}$
regime. For this setup, $N_{a}^{*}$ becomes greater than $1$ for
$\mathsf{SNR}_{0}\approx20$ dB and increases up to the value of $6$
at $\mathsf{SNR}_{0}=50$ dB. Finally, note that the SER performance
curves have the same slope for $N_{a}=1$ and $N_{a}=N_{a}^{*}$ in
the high $\mathsf{SNR}_{0}$ regime, implying that no diversity gain
is achieved from the multipoint transmission \cite{Tse book}. This
is due to the NCJT transmission scheme employed, which can only provide
an $\mathsf{SNR}$ gain \cite{Nigam}.

\section{Conclusion}

The performance of downlink NCJT under practical channel estimation
was analytically characterized under an SG modeling framework that
takes into account randomness of AP positions in dense network deployments.
A worst case assumption in terms of prior information at the UE side
for channel estimation purposes was considered, corresponding to minimal
overhead requirements. The (spatial) average, data detection $\mathsf{SNR}$
was considered as a tractable performance metric, which was characterized
by a simple (semi) closed-form expressions that allows for efficient
evaluation under arbitrary system and operational parameters as well
as (numerical) optimization of important system design parameters
such as length of pilot sequences and number of APs jointly serving
a UE. It was shown that, even though the conventional cellular network
operation with a UE associated only to its closest AP is practically
sufficient under perfect CSI, under practical channel estimation affected
by pilot contamination, coordinated transmissions are optimal with
the achieved gain more pronounced under propagation conditions with
large path loss factors.

An analytical characterization of the minimum pilot sequence length
required to achieve a certain portion of the perfect CSI performance
was obtained, revealing that it is a convex function of the additive
noise level with arbitrarily large overhead required as the noise
level becomes arbitrarily small or large. In addition, an interesting
phase transition phenomenon was observed, where the optimal number
of cooperating APs under arbitrarily large pilot overhead is either
finite or infinite, depending on whether the path loss factor is smaller
or greater than $4$, respectively.

The analysis of this paper is a first step towards understanding the
effects of practical channel estimation on cellular system performance
under the SG modeling framework with various open topics, including
consideration of interference during the data transmission phase and
more sophisticated joint transmission schemes.

\appendices{}

\section{\label{sec:Proof-of-LMMSE estimator formula}Proof of Proposition
\ref{prop: LMMSE estimator formula}}

Since the LMMSE estimator commutes over linear transformations \cite{Kay},
it follows that $\widehat{\mathbf{1}^{T}\mathbf{h}_{\mathcal{C}}}=\mathbf{1}^{T}\hat{\mathbf{h}}_{\mathcal{C}}$,
where $\hat{\mathbf{h}}_{\mathcal{C}}$ is the LMMSE estimate of $\mathbf{h}_{\mathcal{C}}$
based on $\mathbf{y}_{p}$. Towards specifying $\hat{\mathbf{h}}_{\mathcal{C}}$,
note from (\ref{eq: received pilot}) that $\mathbf{y}_{p}$ is a
linear transformation of $\mathbf{h}_{\mathcal{C}}$ due to the application
of the (known at the UE) matrix $\mathbf{P}_{\mathcal{C}}$ plus a
noise term $\tilde{\mathbf{w}}_{p}\triangleq\mathbf{P}_{\bar{\mathcal{C}}}\mathbf{h}_{\bar{\mathcal{C}}}+\mathbf{w}_{p}$.
By assumption, the UE does not know $\mathbf{P}_{\bar{\mathcal{C}}}$,
$\mathbf{h}_{\bar{\mathcal{C}}}$, and $\mathbf{w}_{p}$, and the
AP pilot sequences are independent of the AP channels. Therefore,
the correlation of $\mathbf{h}_{\mathcal{C}}$ with $\tilde{\mathbf{w}}_{p}$
equals 
\begin{align*}
\mathbb{E}(\mathbf{h}_{\mathcal{C}}\tilde{\mathbf{w}}_{p}^{H}) & =\mathbb{E}(\mathbf{h}_{\mathcal{C}}\mathbf{h}_{\bar{\mathcal{C}}}^{H})\mathbb{E}(\mathbf{P}_{\bar{\mathcal{C}}}^{H})+\mathbb{E}(\mathbf{w}_{p}^{H})\\
 & =\mathbf{0},
\end{align*}
since pilot sequences and additive noise are of zero mean. It follows
that the observation $\mathbf{y}_{p}$ coincides with the standard
Bayesian linear model considered in estimation theory and the LMMSE
estimate of $\mathbf{h}_{\mathcal{C}}$ is given as \cite{Kay}
\begin{equation}
\hat{\mathbf{h}}_{\mathcal{C}}=\left(\mathbf{R}_{\mathbf{h}_{\mathcal{C}}}^{-1}+\mathbf{P}_{\mathcal{C}}\mathbf{R}_{\tilde{\mathbf{w}}_{p}}^{-1}\mathbf{P}_{\mathcal{C}}^{H}\right)^{-1}\mathbf{P}_{\mathcal{C}}^{H}\mathbf{R}_{\tilde{\mathbf{w}}_{p}}^{-1}\mathbf{y}_{p},\label{eq:LMMSE estimator, general form}
\end{equation}
where $\mathbf{R}_{\mathbf{h}_{\mathcal{C}}}\triangleq\mathbb{E}(\mathbf{h}_{\mathcal{C}}\mathbf{h}_{\mathcal{C}}^{H})$
and $\mathbf{R}_{\tilde{\mathbf{w}}_{p}}\triangleq\mathbb{E}(\tilde{\mathbf{w}}_{p}\tilde{\mathbf{w}}_{p}^{H})$
are positive definite matrices, as will be verified in the following
where they are explicitly specified in terms of $\sigma_{\mathcal{C}}^{2}$
and $\sigma_{w}^{2}$.

Focusing first on $\mathbf{R}_{\mathbf{h}_{\mathcal{C}}}$, note that
since the UE knows nothing about the AP channels of its serving APs,
the elements of $\mathbf{h}_{\mathcal{C}}$ is a random ordering of
the $N_{a}$ channels $\{h_{\mathbf{x}}\}_{\mathbf{x}\in\mathcal{C}}$.
Therefore, it holds $\mathbf{h}_{\mathcal{C}}=\sum_{\mathbf{x}\in\mathcal{C}}h_{\mathbf{x}}\mathbf{e}_{\mathbf{x}}$,
where $\mathbf{e}_{\mathbf{x}}$ is the unit vector of the standard
basis in $\mathbb{C}^{N_{a}}$ (i.e., all-zeros vector except one
element equal to $1$), with $\mathbf{e}_{\mathbf{x}}\neq\mathbf{e}_{\mathbf{x}'}$
for all $\mathbf{x}\neq\mathbf{x}'\in\mathcal{C}$. Note that, for
any $\mathbf{x}\in\mathcal{C}$, $\mathbf{e_{\mathbf{x}}}$ can be
anyone of the $N_{a}$ basis vectors with equal probability $1/N_{a}$.
Using this representation for $\mathbf{h}_{\mathcal{C}}$, one computes
\begin{align*}
\mathbf{R}_{\mathbf{h}_{\mathcal{C}}} & =\mathbb{E}\left(\sum_{\mathbf{x}\in\mathcal{C}}\sum_{\mathbf{x}'\in\mathcal{C}}h_{\mathbf{x}}h_{\mathbf{x}'}^{*}\mathbf{e}_{\mathbf{x}}\mathbf{e}_{\mathbf{x}'}^{T}\right)\\
 & \overset{(a)}{=}\sum_{\mathbf{x}\in\mathcal{C}}\sum_{\mathbf{x}'\in\mathcal{C}}\mathbb{E}(h_{\mathbf{x}}h_{\mathbf{x}'}^{*})\mathbb{E}(\mathbf{e}_{\mathbf{x}}\mathbf{e}_{\mathbf{x}'}^{T})\\
 & \overset{(b)}{=}\sum_{\mathbf{x}\in\mathcal{C}}\mathbb{E}(|h_{\mathbf{x}}|^{2})\mathbb{E}(\mathbf{e}_{\mathbf{x}}\mathbf{e}_{\mathbf{x}}^{T})\\
 & \overset{(c)}{=}\frac{1}{N_{a}}\sum_{\mathbf{x}\in\mathcal{C}}\mathbb{E}(|h_{\mathbf{x}}|^{2})\mathbf{I}_{N_{a}}
\end{align*}
where $(a)$ is due to the independence of AP channels and AP ordering,
$(b)$ follows since $\mathbb{E}(h_{\mathbf{x}}h_{\mathbf{x}'}^{*})=0$
for any $\mathbf{x}\neq\mathbf{x}'\in\mathcal{C}$, due to the independent
and zero mean fast fading and $(c)$ is obtained by noting $\mathbb{E}(\mathbf{e}_{\mathbf{x}}\mathbf{e}_{\mathbf{x}}^{T})=(1/N_{a})\mathbf{I}_{N_{a}}$,
for all $\mathbf{x}\in\mathcal{C}$. By the independence of the AP
channels, it is easy to see that $\sigma_{\mathcal{C}}^{2}\triangleq\mathbb{E}(|\mathbf{1}^{T}\mathbf{h}_{\mathcal{C}}|^{2})=\sum_{\mathbf{x}\in\mathcal{C}}\mathbb{E}(|h_{\mathbf{x}}|^{2})$,
therefore $\mathbf{R}_{\mathbf{h}_{\mathcal{C}}}$ can be expressed
as
\begin{equation}
\mathbf{R}_{\mathbf{h}_{\mathcal{C}}}=\frac{\sigma_{\mathcal{C}}^{2}}{N_{a}}\mathbf{I}_{N_{a}}.\label{eq: corr_in}
\end{equation}

Turning to $\mathbf{R}_{\tilde{\mathbf{w}}_{p}}$ and writing $\mathbf{P}_{\bar{\mathcal{C}}}\mathbf{h}_{\bar{\mathcal{C}}}$
as $\sum_{\mathbf{x}\in\Phi\setminus\mathcal{C}}h_{\mathbf{x}}\mathbf{p}_{\mathbf{x}}$,
it holds
\begin{align}
\mathbf{R}_{\tilde{\mathbf{w}}_{p}} & =\mathbb{E}\left(\mathbf{P}_{\bar{\mathcal{C}}}\mathbf{h}_{\bar{\mathcal{C}}}\mathbf{h}_{\bar{\mathcal{C}}}^{H}\mathbf{P}_{\bar{\mathcal{C}}}^{H}\right)+\sigma_{w}^{2}\mathbf{I}_{N_{a}}\nonumber \\
 & =\mathbb{E}\left(\sum_{\mathbf{x}\in\Phi\setminus\mathcal{C}}\sum_{\mathbf{x}'\in\Phi\setminus\mathcal{C}}h_{\mathbf{x}}h_{\mathbf{x}'}^{*}\mathbf{p}_{\mathbf{x}}\mathbf{p}_{\mathbf{x}'}^{H}\right)+\sigma_{w}^{2}\mathbf{I}_{N_{a}}\nonumber \\
 & =\sum_{\mathbf{x}\in\Phi\setminus\mathcal{C}}\sum_{\mathbf{x}'\in\Phi\setminus\mathcal{C}}\mathbb{E}(h_{\mathbf{x}}h_{\mathbf{x}'}^{*})\mathbb{E}(\mathbf{p}_{\mathbf{x}}\mathbf{p}_{\mathbf{x}'}^{H})+\sigma_{w}^{2}\mathbf{I}_{N_{a}}\nonumber \\
 & =\sum_{\mathbf{x}\in\Phi\setminus\mathcal{C}}\mathbb{E}(|h_{\mathbf{x}}|^{2})\mathbb{E}(\mathbf{p}_{\mathbf{x}}\mathbf{p}_{\mathbf{x}}^{H})+\sigma_{w}^{2}\mathbf{I}_{N_{a}}\nonumber \\
 & \overset{(a)}{=}\left(\sum_{\mathbf{x}\in\Phi\setminus\mathcal{C}}\mathbb{E}(|h_{\mathbf{x}}|^{2})+\sigma_{w}^{2}\right)\mathbf{I}_{N_{a}}\nonumber \\
 & =\left(\sum_{\mathbf{x}\in\Phi}\mathbb{E}(|h_{\mathbf{x}}|^{2})-\sum_{\mathbf{x}\in\mathcal{\mathcal{C}}}\mathbb{E}(|h_{\mathbf{x}}|^{2})+\sigma_{w}^{2}\right)\mathbf{I}_{N_{a}}\nonumber \\
 & =\left(\sigma_{\Phi}^{2}-\sigma_{\mathcal{C}}^{2}+\sigma_{w}^{2}\right)\mathbf{I}_{N_{a}},\label{eq:corr_out_plus_noise}
\end{align}
where all steps follow by similar arguments as in the derivation of
$\mathbf{R}_{\mathbf{h}_{\mathcal{C}}}$ and $(a)$ is due to $\mathbb{E}(\mathbf{p}_{\mathbf{x}}\mathbf{p}_{\mathbf{x}}^{H})=\mathbf{I}_{N_{a}}$,
for all $\mathbf{x}\in\Phi$, by pilot design assumption. In the last
equality, the fact that $\sigma_{\Phi}^{2}\triangleq\mathbb{E}\left(\left|\sum_{\mathbf{x}\in\Phi}h_{\mathbf{x}}\right|^{2}\right)=\sum_{\mathbf{x}\in\Phi}\mathbb{E}(|h_{\mathbf{x}}|^{2})$,
following by the independence of AP channels, has been used. Its value
can be obtained as 
\begin{align*}
\sigma_{\Phi}^{2} & =\sum_{\mathbf{x}\in\Phi}\mathbb{E}\left(|c_{\mathbf{x}}|^{2}\ell(\|\mathbf{x}\|)\right)\\
 & =\mathbb{E}\left(\sum_{\mathbf{x}\in\Phi}\ell(\|\mathbf{x}\|)\right)\\
 & =\lambda\int_{\mathbf{z}\in\mathbb{R}^{2}}\ell(\|\mathbf{z}\|)d\mathbf{z},
\end{align*}
where the last equality follows by application of Campbel's theorem
\cite{Haenggi Ganti book}. Substituting $\ell(\cdot)$ with its expression
given in (\ref{eq:large scale fading law}) and evaluating the integral
results in the closed form expression of (\ref{eq: total energy}).
The result of the proposition now follows after substituting (\ref{eq: corr_in})
and (\ref{eq:corr_out_plus_noise}) in (\ref{eq:LMMSE estimator, general form})
and some trivial algebra.

\section{\label{sec:proof of average cluster energy}Proof of Proposition
\ref{prop:average cluster energy}}

Let $\rho_{s}>0$ denote the distance from the typical UE of the $s$-th
closest AP whose probability density function equals \cite{Haenggi distances}
\begin{equation}
p(\rho_{s})=\frac{2(\pi\lambda)^{s}}{(s-1)!}\rho_{s}^{2s-1}e^{-\lambda\pi\rho_{s}^{2}},\rho_{s}>0,s\geq1.\label{eq: closest out-of-cluster distance pdf}
\end{equation}
As discussed in the proof of Prop. \ref{prop: LMMSE estimator formula},
the average NCJT channel energy equals $\sigma_{\mathcal{C}}^{2}=\sum_{\mathbf{x}\in\mathcal{C}}\mathbb{E}(|h_{\mathbf{x}}|^{2})$.
It holds
\begin{align}
 & \sum_{\mathbf{x}\in\mathcal{C}}\mathbb{E}(|h_{\mathbf{x}}|^{2})\nonumber \\
= & \sum_{\mathbf{x}\in\mathcal{C}}\mathbb{E}\left(|c_{\mathbf{x}}|^{2}\ell(\|\mathbf{x}\|)\right)\nonumber \\
= & \mathbb{E}\left(\sum_{\mathbf{x}\in\mathcal{C}}\ell(\|\mathbf{x}\|)\right)\nonumber \\
= & \mathbb{E}\left(\mathbb{E}\left(\left.\sum_{\mathbf{x}\in\mathcal{C}}\ell(\|\mathbf{x}\|)\right|\rho_{N_{a}+1}\right)\right)\nonumber \\
\overset{(a)}{=} & \int_{0}^{\infty}p(\rho_{N_{a}+1})\left(N_{a}\int_{0}^{\rho_{N_{a}+1}}\ell(r)\frac{2r}{\rho_{N_{a}+1}^{2}}dr\right)d\rho_{N_{a}+1}\nonumber \\
\overset{(b)}{=} & 2N_{a}\int_{0}^{\infty}\ell(r)r\left(\int_{r}^{\infty}\frac{p(\rho_{N_{a}+1})}{\rho_{N_{a}+1}^{2}}d\rho_{N_{a}+1}\right)dr\nonumber \\
= & 2\pi\lambda\int_{0}^{\infty}\ell(r)r\left(\int_{r}^{\infty}p(\rho_{N_{a}})d\rho_{N_{a}}\right)dr,\label{eq: cluster energy integral}
\end{align}
where $(a)$ follows by fundamental properties of the HPPP, which
state that, conditioned on $\rho_{N_{a}+1}$, the locations of the
$N_{a}$ serving APs are independent and uniformly distributed over
the disk centered at the origin and of radius $\rho_{N_{a}+1}$ \cite{Baccelli},
and $(b)$ by changing the order of integration. Substituting (\ref{eq: closest out-of-cluster distance pdf})
into (\ref{eq: cluster energy integral}) results in (\ref{eq:cluster energy})
after some straightforward algebraic manipulations.

An approximate closed-form expression for (\ref{eq: cluster energy integral})
can be obtained as follows. It can be shown by direct computation
that $\mathbb{E}(\rho_{N_{a}})=\sqrt{N_{a}/(\pi\lambda)}+\mathcal{O}(\sqrt{1/N_{a}}),N_{a}\rightarrow\infty$,
whereas the variance of $\rho_{N_{a}}$ equals $1/(4\pi\lambda)+\mathcal{O}(1/N_{a}),N_{a}\rightarrow\infty$.
This suggests that $p(\rho_{N_{a}})$ is highly concentrated around
its mean for asymptotically large $N_{a}$, suggesting the approximation
$p(\rho_{N_{a}})\approx\delta(\rho_{N_{a}}-\sqrt{N_{a}/(\pi\lambda)})$
in that regime, with $\delta(\cdot)$ denoting the Dirac delta. Substituting
this approximation in (\ref{eq: cluster energy integral}) results
in (\ref{eq: cluster energy approx}).

\section{\label{sec: proof of channel estimation variance}Proof of Proposition
\ref{prop:channel estimation variance}}

Since $\widehat{\mathbf{1}^{T}\mathbf{h}_{\mathcal{C}}}=\mathbf{1}^{T}\hat{\mathbf{h}}_{\mathcal{C}}$,
it holds
\begin{align}
\sigma_{e}^{2} & \triangleq\mathbb{E}(|\mathbf{1}^{T}\mathbf{h}_{\mathcal{C}}-\widehat{\mathbf{1}^{T}\mathbf{h}_{\mathcal{C}}}|^{2})\nonumber \\
 & =\mathbf{1}^{T}\mathbf{R}_{\mathbf{h}_{\mathcal{C}}-\mathbf{\hat{h}}_{\mathcal{C}}}\mathbf{1},\label{eq:var_e general formula}
\end{align}
where $\mathbf{R}_{\mathbf{h}_{\mathcal{C}}-\mathbf{\hat{h}}_{\mathcal{C}}}\triangleq\mathbb{E}\left((\mathbf{h}_{\mathcal{C}}-\mathbf{\hat{h}}_{\mathcal{C}})(\mathbf{h}_{\mathcal{C}}-\mathbf{\hat{h}}_{\mathcal{C}})^{H}\right)$
is the covariance matrix of the zero mean vector $\mathbf{h}_{\mathcal{C}}-\mathbf{\hat{h}}_{\mathcal{C}}$.
With $\mathbf{R}_{\mathbf{h}_{\mathcal{C}}}$, $\mathbf{R}_{\tilde{\mathbf{w}}_{p}}$
as defined in Appendix \ref{sec:Proof-of-LMMSE estimator formula},
$\mathbf{R}_{\mathbf{h}_{\mathcal{C}}-\mathbf{\hat{h}}_{\mathcal{C}}}$
equals \cite{Kay}
\begin{align}
\mathbf{R}_{\mathbf{h}_{\mathcal{C}}-\mathbf{\hat{h}}_{\mathcal{C}}} & =\left(\mathbf{R}_{\mathbf{h}_{\mathcal{C}}}^{-1}+\mathbf{P}_{\mathcal{C}}^{H}\mathbf{R}_{\tilde{\mathbf{w}}_{p}}^{-1}\mathbf{P}_{\mathcal{C}}\right)^{-1}\nonumber \\
 & =\left(\frac{N_{a}}{\sigma_{\mathcal{C}}^{2}}\mathbf{I}_{N_{a}}+\frac{1}{\sigma_{w}^{2}+\sigma_{\Phi}^{2}-\sigma_{\mathcal{C}}^{2}}\mathbf{P}_{\mathcal{C}}^{H}\mathbf{P}_{\mathcal{C}}\right)^{-1},\label{eq:chan estimation error covariance matrix}
\end{align}
where the second equality follows from (\ref{eq: corr_in}) and (\ref{eq:corr_out_plus_noise}).
Substituting (\ref{eq:chan estimation error covariance matrix}) into
(\ref{eq:var_e general formula}) gives $\sigma_{e}^{2}$, which,
as expected, is a function of the AP cluster pilot sequences contained
in $\mathbf{P}_{\mathcal{C}}$. However, a simpler approximate expression,
independent of $\mathbf{P}_{\mathcal{C}}$ can be obtained as follows.
With increasing $N_{p}$, $\mathbf{p}_{\mathbf{x}}^{H}\mathbf{p}_{\mathbf{x}'}\rightarrow0$,
for all $\mathbf{x}\neq\mathbf{x}'\in\mathcal{C},$ suggesting that
$\mathbf{P}_{\mathcal{C}}^{H}\mathbf{P}_{\mathcal{C}}$ is approximately
diagonal for large $N_{p}$. This, in turn, implies that $\mathbf{R}_{\mathbf{h}_{\mathcal{C}}-\mathbf{\hat{h}}_{\mathcal{C}}}$
is approximately diagonal for large $N_{p}$ and taking this into
account in (\ref{eq:var_e general formula}) results in 
\begin{align*}
\sigma_{e}^{2} & \approx\text{trace}\left(\left(\frac{N_{a}}{\sigma_{\mathcal{C}}^{2}}\mathbf{I}_{N_{a}}+\frac{1}{\sigma_{w}^{2}+\sigma_{\Phi}^{2}-\sigma_{\mathcal{C}}^{2}}\mathbf{P}_{\mathcal{C}}^{H}\mathbf{P}_{\mathcal{C}}\right)^{-1}\right).
\end{align*}
For $N_{a},N_{p}\rightarrow\infty$, with the ratio $N_{a}/N_{p}$
constant, this expression converges to the limit given in (\ref{eq:error variance})
for any realization of the AP pilot sequences \cite[Eq. (1.16)]{Tulino RMT}.

\section{\label{sec:Proof of minimum pilot length proposition}Proof of Proposition
\ref{prop:Minimum pilot length}}

Treating $N_{p}$ as a real number, its minimum required value is
obtained by solving the non-linear equation $\mathsf{SNR}=\gamma\sigma_{\mathcal{C}}^{2}/\sigma_{w}^{2}$
w.r.t. $N_{p}$. Assume that the minimum solution $N_{p}^{*}$ of
this equation is much greater than $1$. Then, this $N_{p}^{*}$ can
be obtained approximately by replacing $\mathsf{SNR}$ in the above
equation with its expression given in (\ref{eq:SNR expression with LMMSE chan. est.})
using the large-$N_{p}$ approximation of $\sigma_{e}^{2}$ given
in (\ref{eq:error variance}). Solving the resulting equation w.r.t.
$N_{p}$ gives the solution of (\ref{eq:N_p formula}). However, the
formula of (\ref{eq:N_p formula}) should be used with caution since,
for a given set of system parameters appearing in the right hand side
of (\ref{eq:N_p formula}), it may happen that the approximate $N_{p}^{*}$
is small or even negative, which contradicts the assumption of large
$N_{p}^{*}$ based on which (\ref{eq:N_p formula}) was derived. Therefore,
conditions resulting in a large approximate $N_{p}^{*}$ as per (\ref{eq:N_p formula})
that is a good approximation of the true (and large) $N_{p}^{*}$
must be found.

One approach to identify such conditions is to obtain a lower bound
for the right-hand side of (\ref{eq:N_p formula}) and identify the
conditions which guarantee that this bound is large. The resulting
conditions will clearly be sufficient although they may not be necessary.
Let $g$ denote the right-hand side expression of (\ref{eq:N_p formula}).
It is easy to see that $g$ is a convex function of $\sigma_{w}^{2}$
with $\min_{\sigma_{w}^{2}}g=\frac{\gamma N_{a}\sigma_{\Phi}^{2}}{(1-\gamma)\sigma_{\mathcal{C}}^{2}}$,
achieved at $\sigma_{w}^{2}=\sigma_{\mathcal{C}}(\sigma_{\Phi}-\sigma_{\mathcal{C}})$.
Noting that it holds $\sigma_{\mathcal{C}}^{2}\leq\sigma_{\Phi}^{2}$,
it follows that $\min_{\sigma_{\mathcal{C}}^{2},\sigma_{w}^{2}}g=\frac{\gamma N_{a}}{(1-\gamma)}$
resulting in the sufficient condition of (\ref{eq:N_p condition}).

\section{\label{sec:Proof of optimal cluster size}Proof of Proposition \ref{prop: optimal cluster size}}

By setting $\sigma_{w}^{2}=0$ in (\ref{eq:SNR expression with LMMSE chan. est.})
and using (\ref{eq:error variance}), it holds 
\begin{align}
\mathsf{SNR} & =\frac{N_{p}}{N_{a}}\left(\frac{\frac{\sigma_{\Phi}^{2}-\sigma_{\mathcal{C}}^{2}}{\sigma_{\mathcal{C}}^{2}}}{f_{N_{a}/N_{p}}\left(\frac{\sigma_{\Phi}^{2}-\sigma_{\mathcal{C}}^{2}}{\sigma_{\mathcal{C}}^{2}}\right)}\right)-1\nonumber \\
 & \overset{(a)}{=}\left(\frac{N_{p}}{N_{a}}-1\right)\frac{\sigma_{\mathcal{C}}^{2}}{\sigma_{\Phi}^{2}-\sigma_{\mathcal{C}}^{2}}+\mathcal{O}\left(\left(\frac{N_{a}}{N_{p}}\right)^{2}\right)\nonumber \\
 & =\frac{N_{p}\sigma_{\mathcal{C}}^{2}}{N_{a}(\sigma_{\Phi}^{2}-\sigma_{\mathcal{C}}^{2})}\left(1-\mathcal{O}\left(\frac{N_{a}}{N_{p}}\right)\right),\frac{N_{a}}{N_{p}}\rightarrow0,\label{eq:SNR large N_p}
\end{align}
where $(a)$ follows by noting that $b/f_{a}(b)=b+a(1-b)+\mathcal{O}(a^{2}),a\rightarrow0$.
For $N_{a}\gg1$, the approximation of (\ref{eq: cluster energy approx})
for $\sigma_{\mathcal{C}}^{2}$ can be employed in (\ref{eq:SNR large N_p})
resulting in 
\[
\mathsf{SNR}=\frac{N_{p}}{N_{a}}\left[\frac{\alpha}{2}\left(\frac{N_{a}}{\lambda\pi r_{0}^{2}}\right)^{\frac{\alpha}{2}-1}-1\right]\left(1-\mathcal{O}\left(\frac{N_{a}}{N_{p}}\right)\right).
\]
It is easy to see that for $\alpha>4,$ $\mathsf{SNR}$ is increasing
with increasing $N_{a}$ (assuming always that $N_{p}\gg N_{a})$,
therefore, an arbitrarily large $N_{a}$ is $\mathsf{SNR}$-optimal
in this regime, whereas $\mathsf{SNR}$ is decreasing with increasing
$N_{a}$ for $\alpha\leq4$, thus the $\mathsf{SNR}$-optimal $N_{a}$
must be finite in this regime.

\end{document}